\documentclass[a4paper]{article}
\usepackage[left=2cm,top=2cm,bottom=2cm,right=2cm,nohead,nofoot]{geometry}
\pdfoutput=1

\usepackage{lineno,hyperref}
\usepackage[affil-it]{authblk}
\usepackage{booktabs} 
\usepackage{graphicx}
\usepackage{threeparttable}
\usepackage{arydshln}
\usepackage{caption}
\usepackage[numbers]{natbib}
\graphicspath{{AFigs/}}

\title{Detecting driver distraction using stimuli-response EEG analysis}

\author[1,2]{Garima Bajwa\thanks{corresponding author}}
\author[1]{Mohamed Fazeen}
\author[1]{Ram Dantu}

\affil[1]{University of North Texas, 1155 Union Circle, Denton, TX, USA}
\affil[2]{Capitol Technology University, 11301 Springfield Rd, Laurel, MD, USA}

\begin{document}
\maketitle

\begin{abstract}
\textbf{Introduction}: Detecting driver distraction is a significant concern for future intelligent transportation systems. We present a new approach for identifying a distracted driver's behavior by evaluating a stimulus and response interaction with the brain signals in two ways. \emph{First, measuring driver's response through EEG by creating various types of distraction stimuli such as reading, texting, calling and using phone camera (risk odds ratio of these activities determined by NHTSA study). Second, using a survey, comparing driver's order/perception of severity of distraction with the derived distraction index from EEG bands.}

\textbf{Method}: A 14 electrodes headset was used to record the brain signals while driving in the pilot study with two subjects and a single dry electrode headset with 13 subjects in the main study. We used a naturalistic on-the-road driving study as opposed to a virtual-reality driving simulator to perform the distracted driving maneuvers, consisting of over 100 short duration trials (three to five seconds) for a subject.

\textbf{Results}: We overcame a big challenge in EEG analysis \textendash reducing the number of electrodes by isolating one electrode (FC5) from 14 electrode locations to identify certain distractions. Our machine learning methods achieved a mean accuracy (averaged over the subjects and tasks) of 91.54 $\pm$ 5.23\% to detect a distracted driving event and 76.99 $\pm$ 8.63\% to distinguish between the five distraction cases in our study (read, text, call, and snapshot) using a single electrode.


\textbf{Practical Applications}: The quantification of distracted driving detailed in this paper is necessary to guide future policies in road safety. Our real\textendash time system addresses the safety concerns resulting from driver distraction and aims to bring about behavioral changes in drivers.

\end{abstract}


\section{Background}
Electroencephalography (EEG) is a technique for the measurement of electrical activity or the potential difference between different parts of the brain. It occurs as a result of ionic current flows when neurons communicate with each other. The novel idea of using EEG to communicate with a computer emerged in the early 70's with the research initiated by Vidal \cite{annurev}. Since then, many diverse areas such as robotics, gaming, and neurofeedback applications use this technique \cite{SSS124355}. Brain-computer-interfaces (BCI) are being adopted increasingly for real-time monitoring of patients, neural prosthetics, affective computing, gaming, and security \cite{Wolpaw2002767, muhl2014survey, bajwa2016neurokey}. The future developments are being focused on recording and transmitting EEG signals using a wireless platform and at the same time supporting mobility for such applications \cite{lin2009review, campbell2010neurophone, wang2011cell}.

In the recent past, many investigations have been made for driver distraction behavior. Specifically, the U.S. National Highway Traffic Safety Administration (NHTSA) has given significant attention to this issue \cite{6, 1, 2}. It reported nearly 3,450 deaths in 2016 (9\% of overall fatalities) resulting from distracted driving crashes in the United States \cite{nhtsa2016}. NHTSA defines driver distraction as \emph{``any activity that takes a driver's attention away from the task of driving"}. Their guidelines \cite{7} have listed manual text entry, reading and displaying graphics unrelated to driving as unsafe driver distractions. According to studies on drivers' willingness to engage in various sorts of potentially distracting tasks, highest mean ratings were shown in activities like entering text messages from the phone, looking up stored phone numbers, picking up and reading a message on PDA and sending emails \cite{3, caird2014use}. It influenced us to formulate an experiment to study an individual's brain pattern involved in common distraction activities.

Measuring whether a driver is distracted or not is not an easy task and may involve analysis of many parameters \cite{carney2018examining}. There must be proper metrics and measurements available to quantify driver distraction accurately. Obtaining data directly from the brain to measure distraction sounds more natural and plausible. According to a review by Dong et al. \cite{5164395}, there are primarily three types of inattention detection methods \textendash biological signal processing approaches, subjective report approaches and behavior signal processing approaches. Under biological signal processing approaches, EEG plays a critical role in detecting driver distractions. Further, their study highlights EEG technique as one of the most accurate methods of detecting driver inattention. Their findings inspired the usage of EEG in our research to study about driver distraction.

\begin{figure}[h]
\centering
  \includegraphics[width=0.6\linewidth]{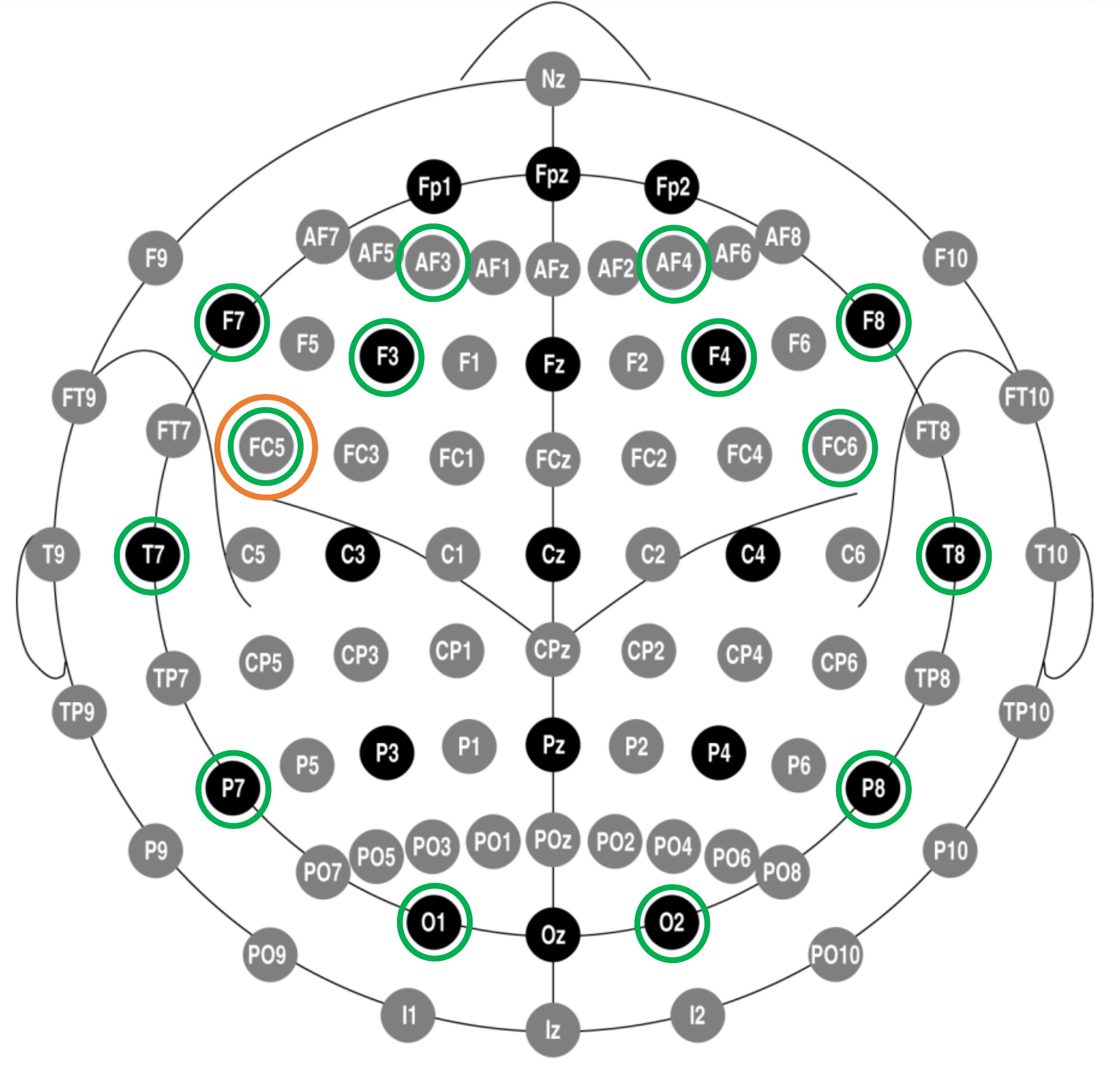}
    \caption{\textbf{10-10 Electrode positioning system.} Black circles indicate positions of the original 10-20 system, gray circles indicate additional positions introduced in the 10-10 extension \cite{oostenveld2001five}. The letters F, T, C, P and O stand for frontal, temporal, central, parietal, and occipital lobes, respectively. The electrodes used in EPOC headset are marked in green color and Neurosky electrode placement in orange.}
  \label{fig:elecs}
\end{figure}

Most of the existing work for analyzing driver distraction using EEG was performed in simulated environments \cite{almahasneh2014deep, 5479148, schier2000changes, lal2003development, 6033401}. Faro et al. \cite{4463199} in their study of analyzing driver status observed that channel locations F7, F8, FC5 (according to international 10$-$20 system, Fig. \ref{fig:elecs}) play an active role in detecting distraction. They observed the frontal areas as the most significant channels for 18 out of 20 peoples' sample. Using a virtual-reality (VR) based simulator, Lin et al. \cite{lin2011spatial} observed increased powers in the theta, and beta frequency bands in the frontal cortex while studying the dynamics of dual-task driving performance. They also proposed that motor area was not related to the distraction effects as most of the brain resources were occupied in the frontal area to deal with the two tasks. Also, the correlation was low between EEG dynamics in motor area and its corresponding response times.

Detecting driver emotions using EEG can also be very useful in observing driver distraction. Fan et al. \cite{5580919} explained the importance of emotions towards driver safety. Their research claims to detect human emotion using EEG with an accuracy of 72.25\% for the targeted group samples. Studies in the past have also evaluated drivers' mental workload \cite{4706518}. Lei, Welke and Roetting \cite{4} assessed this workload using EEG data when drivers were performing multiple tasks simultaneously. They also used a virtual driving simulator to present each subject with four different blocks of lane change tasks and observed that EEG is very effective tool for evaluating a driver's mental workload. An EEG-based drowsiness estimation system used the EEG power spectrum methods to estimate indirectly a driver's drowsiness level in a VR-based driving simulator \cite{lin2005estimating}. Another work by Lin et al. showed the use of independent component analysis (ICA) in estimating driver's drowsiness effectively \cite{1693037}.
\subsection{\textbf{Contributions}}
Current literature focuses on the use of simulated driving in distraction studies using EEG.  In this work, we collected responses from human subjects during on-the-road driving in a real environment that incorporates complexities arising from multisensory cues as opposed to virtual\textendash reality driving that involves only distal cues and non-vestibular self-motion cues  \cite{perani2001different, han2005distinct, ravassard2013multisensory}. The newer generation of EEG recording devices \cite{Neurosky, Emotiv} have eased the signal acquisition process and storage. Our aim wad to utilize these devices to develop a reliable and robust application that supports mobility as well as simplification of high-dimensional EEG data to generate safe real-time feedbacks for the drivers'. A user acceptable assistive application is highly desirable to disrupt and change behavior, and prevent fatal accidents resulting from distracted driving.\\

The key contributions of this paper are:
\begin{itemize}
   \item Performed extensive on-the-road trials with 15 subjects in a real-life naturalistic driving setup where the brain responses are much more complex and intriguing than a simulated driving environment. Nearly all the prior studies focused on the use of simulated study  \cite{lin2011spatial, wang2014eeg, wang2015eeg}.
    \item Localized the identification of common distractions to one electrode (FC5) from 14 electrode locations. All the previous work performed classification and prediction with more than one electrode \cite{5164395}. Using a single dry sensor EEG also increases the usability and acceptability of our system.
   \item Established a system to understand and distinguish between multiple scenarios of everyday driver distractions such as reading, texting, calling and taking a snapshot.
   \item Developed an interactive real-time training and calibration of a mobile application to provide safety alerts to the drivers based on their level of cognitive distraction computed using the EEG spectrum.
\end{itemize}

\begin{figure}
\centering
  \includegraphics[width=0.8\linewidth]{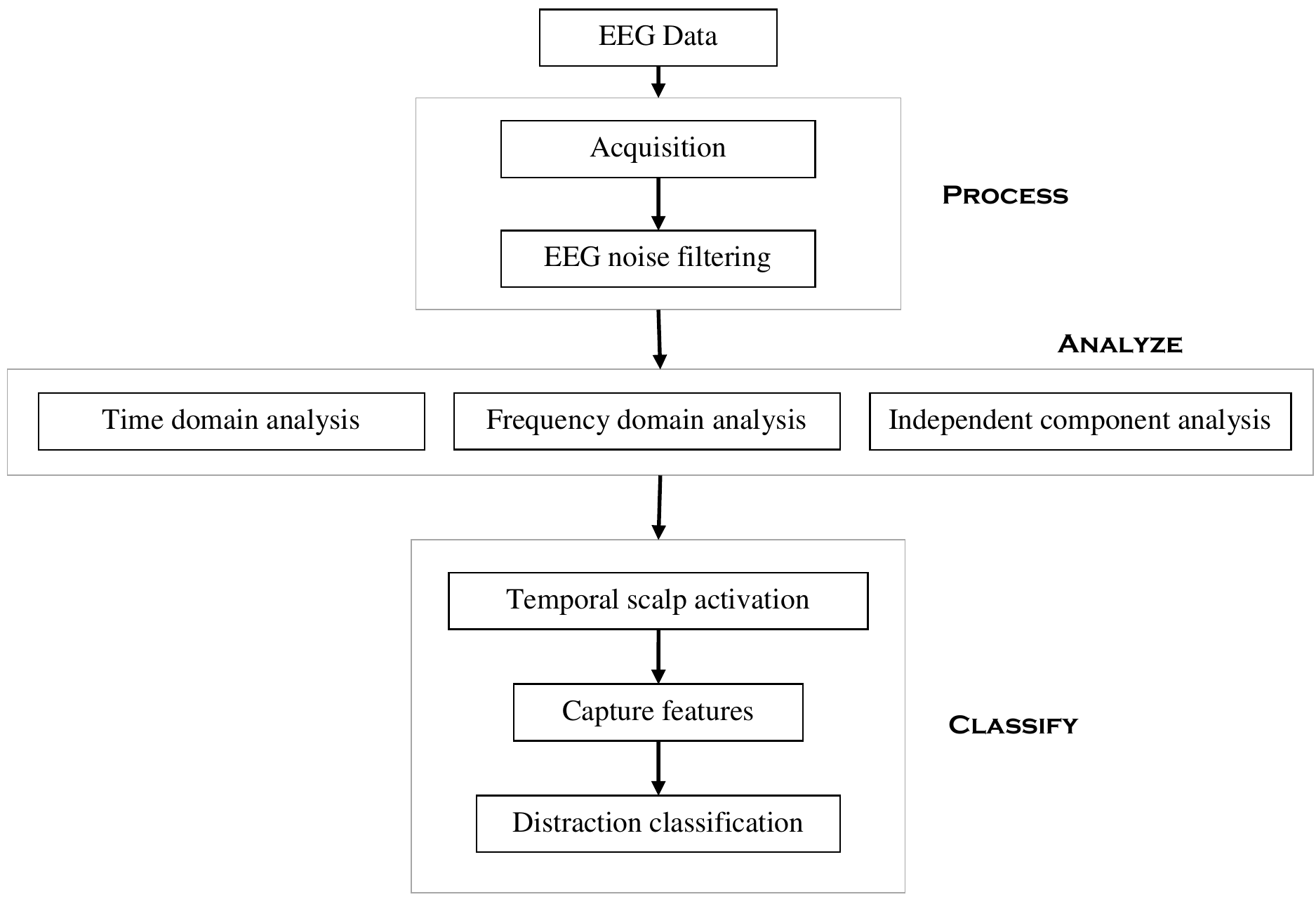}
     \caption{\textbf{A schematic depiction of the processing stages of the distraction detection system.} The first process stage deals with EEG signal acquisition and filtering. In the analyze stage, three techniques \textendash time, frequency and independent components were used to understand the signals behavior in distractions. The classify stage incorporates feature formation, selection, and classification evaluation of various distraction events.}
  \label{fig:sys}
\end{figure}

The methodology of the driver distraction system consists of three main stages; \emph{process} the EEG signals, \emph{analyze} using time, frequency and independent components, and \emph{classify} them into various types of distraction events (Fig. \ref{fig:sys}). The rest of the paper discusses scenarios of distraction and the methods that help to recognize such behaviors. The last section is dedicated to the feasibility of detecting these events on a mobile platform.

\section{Materials and Methods}
\subsection{\textbf{Ethics Statement}}
Ethical approval for the study was obtained from the Institutional Review Board at the University where research was conducted.
\subsection{Stimuli}
\begin{figure}
\centering
  \includegraphics[width=0.9\linewidth]{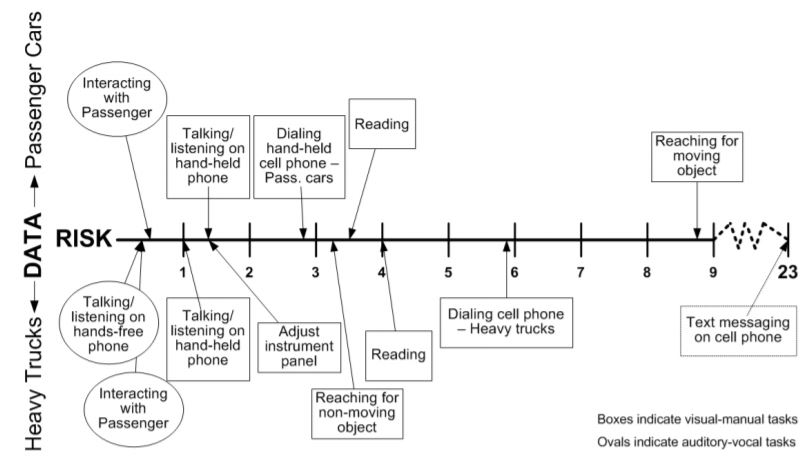}
  \caption{The risk odds ratio determined by the NHTSA study \cite{national2012visual}}
  \label{fig:nhtsa}
\end{figure}

The stimuli used in this study were based on the common distractions (read, text, call, $\&$ snapshot) observed while driving \cite{1, klauer2006impact, national2012visual}. The NHTSA study \cite{national2012visual} provides a summary of the risks associated with performing these distraction tasks (Fig. \ref{fig:nhtsa}). Following is a brief description of all the stimuli used to study the EEG responses:
\begin{enumerate}
\item \emph{Base (undistracted)}: Normal driving behavior of a subject without engaging in any tasks while driving
\item \emph{Read}: Subject was presented with an unseen article on a printed paper to read while driving
\item \emph{Text}: Involved composing and sending a message while driving by using any of the standard messaging services on the mobile phone
\item \emph{Call}: Subject browsed his/her contact list and made a call to the desired person
\item \emph{Snapshot}: Subject operated the phone's camera that involved taking pictures either using the front or back camera
\end{enumerate}

We also performed a pretending to text activity to observe the differences from an actual distraction task. In this activity, the driver looked at his phone screen intermittently while driving. The testing conditions were purposely kept simple in order to implement a controlled driving environment by isolating only the designated distraction. Adding complicated driving courses and behaviors such as hard turns or lane changes cause more complex brain activity that could mask or mislead the driver distraction signals.

\subsection{\textbf{Driving Location}}
\begin{figure}
\centering
  \includegraphics[width=0.6\linewidth]{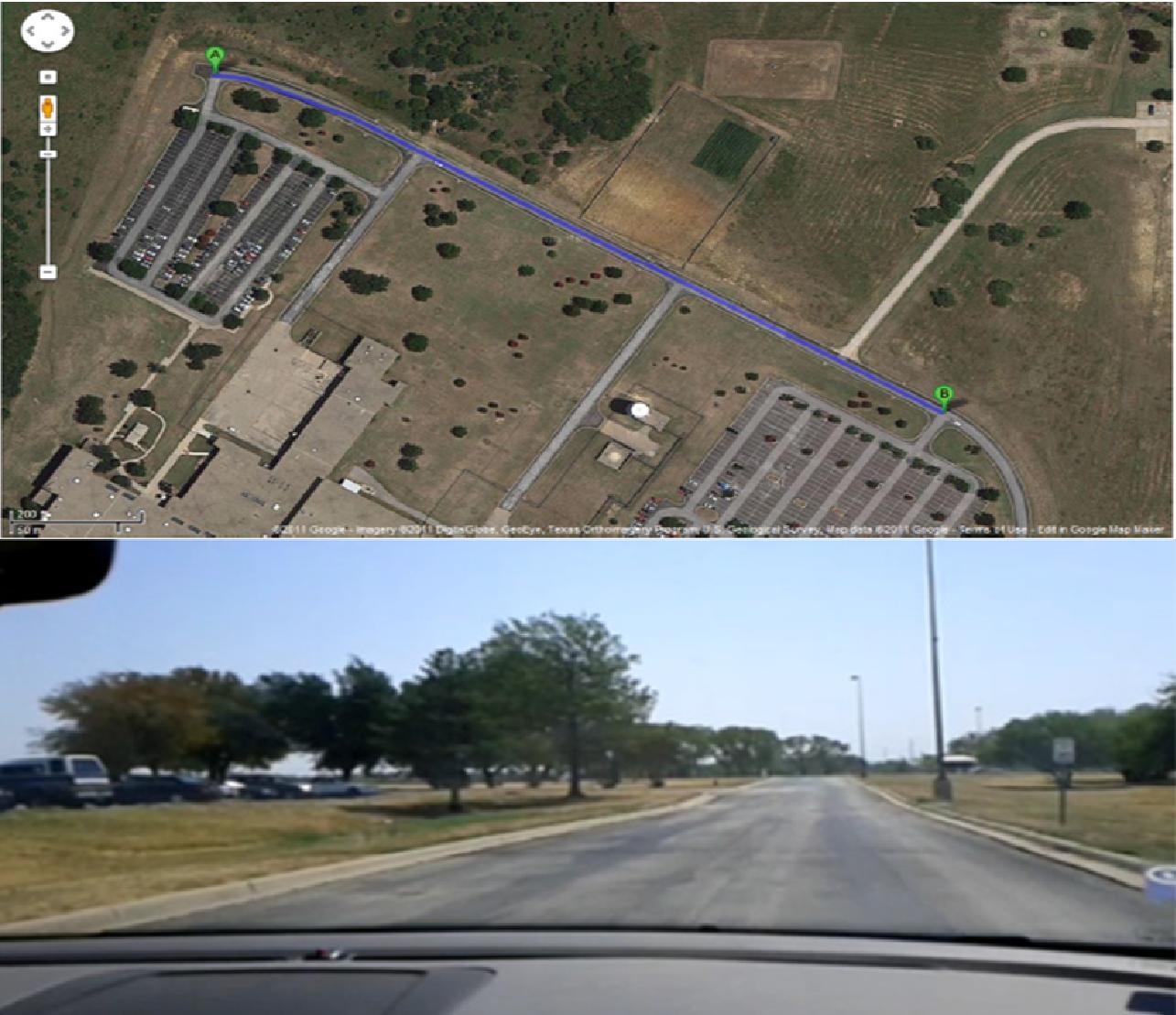}
  \caption{\textbf{A view of the road section inside the university campus used to conduct the distracted driving experiment.}}
  \label{fig:road}
\end{figure}

 The road section adjacent to a parking lot on the University campus was chosen for the study (Fig. \ref{fig:road}). We conducted the experiment during the time the parking lot was rarely used to ensure a safe environment for the study. The path driven was a straight and curved stretch of nearly one mile. There was a minor risk of accident associated with performing the tasks while driving. Therefore, principal investigator and the student investigator constantly kept a check for any undesirable events such as subjects moving out of the lane while driving, monitoring incoming traffic on the opposite lane, and rare cases of any pedestrians being present to make sure of the safety on the road. There was no case of any injury or accident at the end of data collection process from any subject.

\subsection{\textbf{EEG Acquisition Devices}}
The distracted driving study was conducted using two different EEG recording devices; Emotiv's EPOC neuroheadset \cite{Emotiv} and Neurosky's Mindband \cite{Neurosky}. EPOC is a portable device to record EEG data from 14 saline electrodes placed according to the international 10$-$20 system (Fig. \ref{fig:elecs}). Its sampling rate is 128 Hz. It has a built-in gyroscope that generates optimal positional information for cursor and camera controls allowing a total range of motion. The Neurosky headband contains a single sensor dry electrode with an ear clip reference to record signals from the scalp. Its sampling rate is 512 Hz and records signals up to 100Hz. It has a rechargeable battery, and a stretchable fabric makes it very convenient to wear and adjustable to different head sizes. A significant benefit over Emotiv is that it has a single dry electrode and needs no preparation. Since, the data collection was carried out in a dynamic environment with multiple subjects, we used conductive gel in some cases to ensure good contact with the Neurosky electrode. Quality of the signals was monitored using the signal strength information from the electrode (good strength equivalent to 200 is solid contact with the skin/scalp).
\subsection{\textbf{Data Collection}}
The data was recorded for 15 right handed subjects, nine males, and five females with ages between 24 \textendash 28 years, and one male with age 55 years. In the pilot study with two subjects, EEG data was recorded from both the 14 electrodes and single electrode headsets. In the main study with 13 subjects, EEG data was collected using the single electrode device (explained in forthcoming sections). A session consisted of performing one trial of each distraction activity (read, text, call, and snapshot). The recording time for each task was nearly 60 seconds over the pre-defined road route. We avoided contaminating the EEG readings of participants with any known factors such as tiredness or boredom. At the end of every session, the participants either chose to come back or continue with the data collection. The variation in sessions was dependent upon the willingness of participants to volunteer for more than one session. Hence, the number of sessions varied from two to four for the subjects. The details of the complete dataset are shown in Table \ref{tab:data}.

\begin{table*} [h!]
\captionsetup{width=\linewidth}
\caption{Statistics of the dataset. Each row corresponds to the details of the trials for a subject performing the various distraction tasks. Duration of each trial varies from three to five seconds.} 
  \centering
  \begin{threeparttable}[b]
   \scalebox{0.8}{
\begin{tabular}{|c|c|*5c|c|} \hline\hline
Subject & Interface  & Baseline	& 	Text	& 	Read	& 	Snapshot	& 	Call	& 	Total \\
	& 	\multicolumn{6}{|c|}{Number of trials}	\\	\hline
1	&	BCI	&	20	&	20	&	20	&	20	&	20	&	100	\\	
2	&	BCI	&	20	&	20	&	20	&	20	&	20	&	100	\\	\hline
3	&	BCI	&	25	&	25	&	25	&	25	&	25	&	125	\\	
4	&	BCI	&	25	&	25	&	25	&	25	&	25	&	125	\\	
5	&	BCI	&	15	&	15	&	15	&	15	&	15	&	75	\\	
6	&	BCI	&	25	&	25	&	25	&	25	&	25	&	125	\\	
7	&	BCI	&	30	&	30	&	30	&	30	&	30	&	150	\\	
8	&	BCI	&	30	&	30	&	30	&	30	&	30	&	150	\\	
9	&	BCI	&	30	&	30	&	30	&	30	&	30	&	150	\\	
10	&	BCI	&	25	&	25	&	25	&	25	&	25	&	125	\\	
11	&	BCI	&	15	&	15	&	15	&	15	&	15	&	75	\\	
12	&	BCI	&	25	&	25	&	25	&	25	&	25	&	125	\\	
13	&	BCI	&	35	&	35	&	35	&	35	&	35	&	175	\\	
14	&	BCI	&	30	&	30	&	30	&	30	&	30	&	150	\\	\hline
15	&	BMI	&	60	&	69	&	69	&	69	&	69	&	336 \\	\hline	\hline

\end{tabular}
}
  \begin{tablenotes}
   \item BCI - Brain-Computer-Interface, BMI - Brain-Mobile-Interface
  \end{tablenotes}
 \end{threeparttable}
  \label{tab:data}
\end{table*}

The experiment was carried out by two people; one performed the driving task, and the other person recorded the data as well as gave the audio prompts for the desired activity. The driver was asked to relax for few seconds before a prompt to start driving. An audio cue was given to perform a certain distraction activity (text or read) and finally, the driver was asked to stop the vehicle safely. The subject was not aware of the order in which he/she will perform the distraction tasks. We wanted to carry the distraction activity as a controlled event to visualize the maximum effect of distraction in the EEG activity. Also, we did not want to bias the EEG signals with any known criteria prior to performing the distraction activity and intended to keep the experimental conditions similar for all participants.

A laptop was used to establish a wireless connection to the EEG headsets and record the data for further processing. A video of the driver was also recorded simultaneously throughout the experiment for reference and verification of the distraction tasks. Both the EEG and video recording were synchronized at all times. The same EEG data was also collected synchronously on the smartphone using the mobile API (Application Interface) described in the section on Brain Mobile Application Interface. During the whole experiment, the second person in the vehicle also kept a check on any undesirable events. In such a situation, the experiment was aborted and started again.

\section{Stimuli-Response Identification}
The challenge in this experiment was whether we could localize the identification of response to distracted stimuli using a few scalp locations. The data collected from 14 electrodes added many parameters and constraints for carrying out effective real-time analysis on the recorded brain signals. Also, the subjects preferred to wear an EEG cap with less number of electrodes. So, the initial phase of the experiment involved reducing the electrode positions to study the EEG responses to stimuli with a 14 electrode headset. We performed three analysis - \emph{\textbf{time domain, frequency domain and independent component analysis}} to discover the electrodes capable of detecting distraction with reasonable accuracy.
\begin{figure}
\centering
  \includegraphics[width=0.8\linewidth]{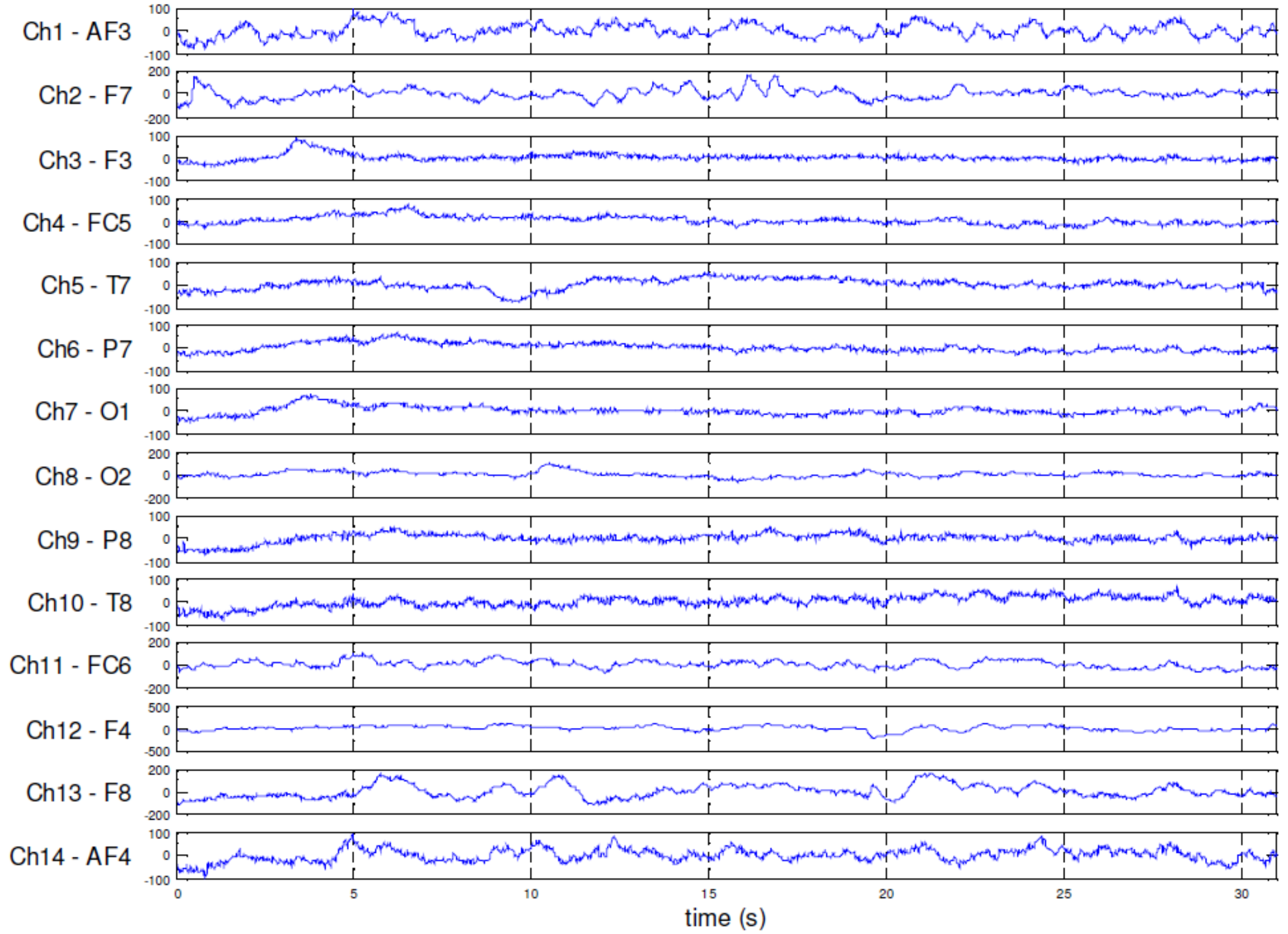}
  \caption{\textbf{EEG signals from a base profile (undistracted driving) trial of a subject using the 14 electrodes headset.}}
  \label{fig:base}
\end{figure}

\subsection{\textbf{Time Domain Analysis}}
\begin{figure}[htb]
\centering
  \includegraphics[width=0.8\linewidth]{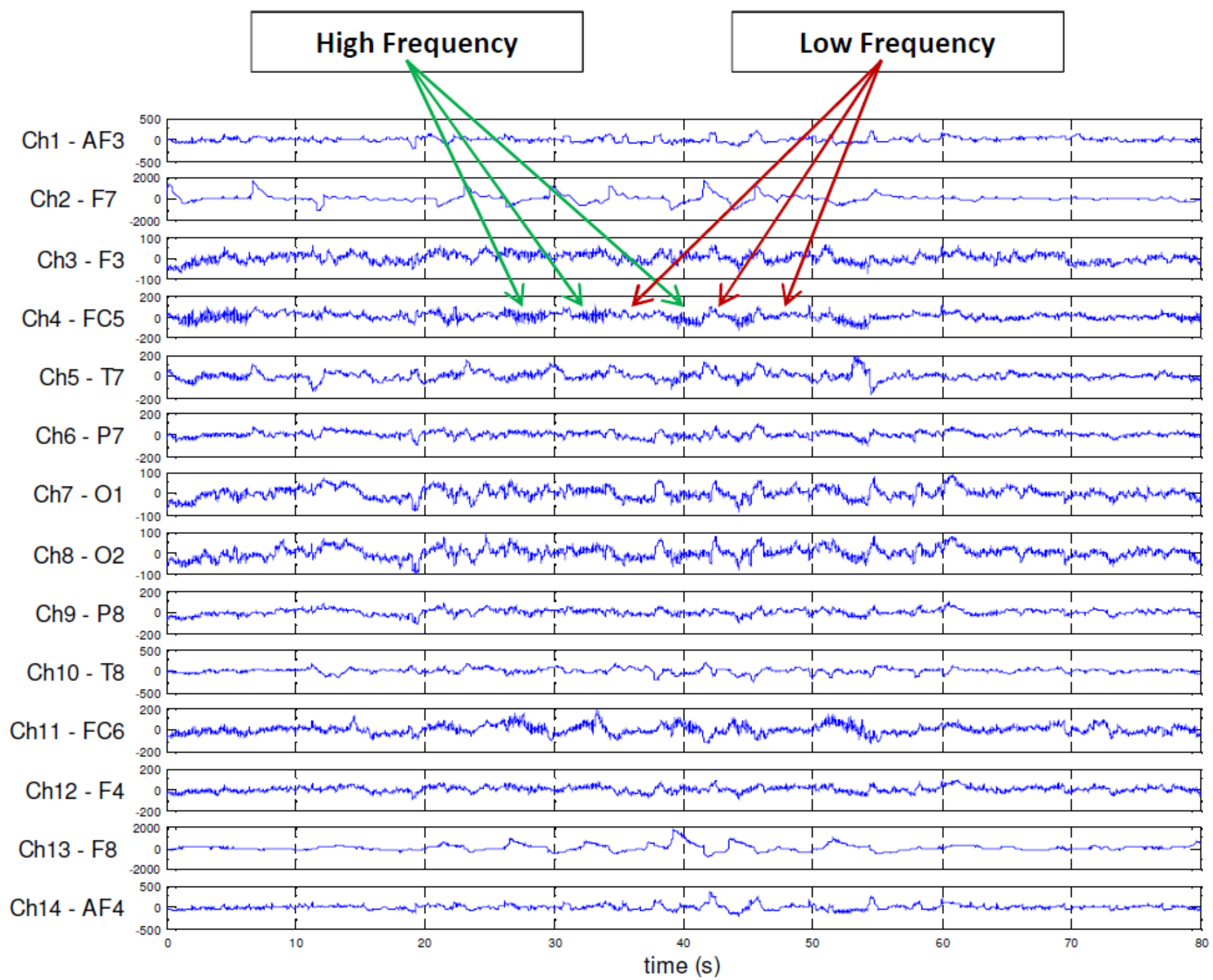}
  \caption{\textbf{EEG signals from a texting while driving trial of a subject using the 14 electrodes headset.} Signals corresponding to the O1, O2, FC5 and FC6 electrode regions exhibit alternating high-low frequency patterns in the raw EEG compared to the baseline signals (Fig. \ref{fig:base}).}
  \label{fig:text}
\end{figure}

A base trial (undistracted activity) was recorded before each distracted driving task trial using 14 electrodes. This collection generated a large number of data readings. Therefore, for illustration purpose, we plot the brain activity of only one trial of two activity types for a subject. Figs. \ref{fig:base} and \ref{fig:text} show the brain activity during normal driving and sending a text message while driving. O1, O2, FC5 and FC6 electrodes showed a rhythmic pattern of occurrence of high and low frequency in the EEG signals of text activity compared to the baseline signals (statistical tests detailed in Results and Discussion section).
Shweizer et al. fMRI study suggested that brain activation shifted dramatically from the posterior, visual, and spatial areas to the prefrontal cortex during distracted driving \cite{schweizer2013brain}. We also observed that channel-4 (FC5) showed sudden bursts of high and low frequency in the brain signals compared to other electrodes for various types of distraction events.
\begin{figure} [h!]
\centering
  \includegraphics[width=0.7\linewidth]{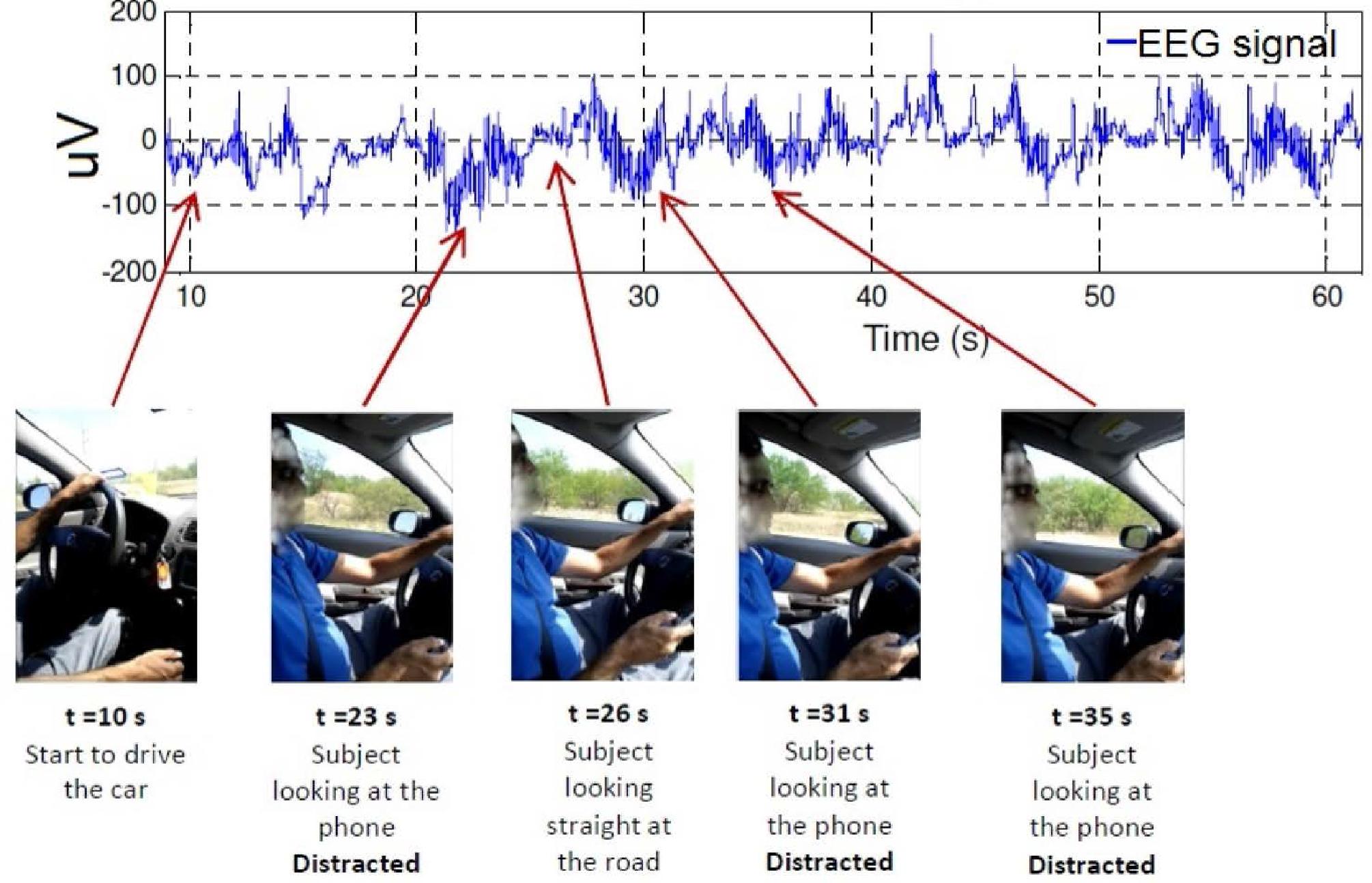}
  \caption{\textbf{An example of the time frames extracted from the video recording of a texting while driving activity.} These recordings were used to verify the observed changes in the EEG signals corresponding to the distraction events captured in the video frames.}
  \label{fig:video}
\end{figure}
We further investigated to verify if the EEG behavior identified in the FC5 channel correlated to driver's distracted activity. As the recorded video of the driver was synchronized with the EEG data recording, it was easy to find a correlation between our observations and the behavior exhibited by the EEG signals. Fig. \ref{fig:video} shows the moments of distraction extracted from the video frames of the distracted activity. The time of the extracted event shown under each frame explains the corresponding EEG pattern. The text while driving activity showed the appearance of rhythmic occurrence of EEG bursts corresponding to the action of typing a text while looking at the screen of the phone.
%
\subsection{\textbf{Time-Frequency Analysis}}
\begin{figure}[h!]
\centering
  \includegraphics[width=0.7\linewidth]{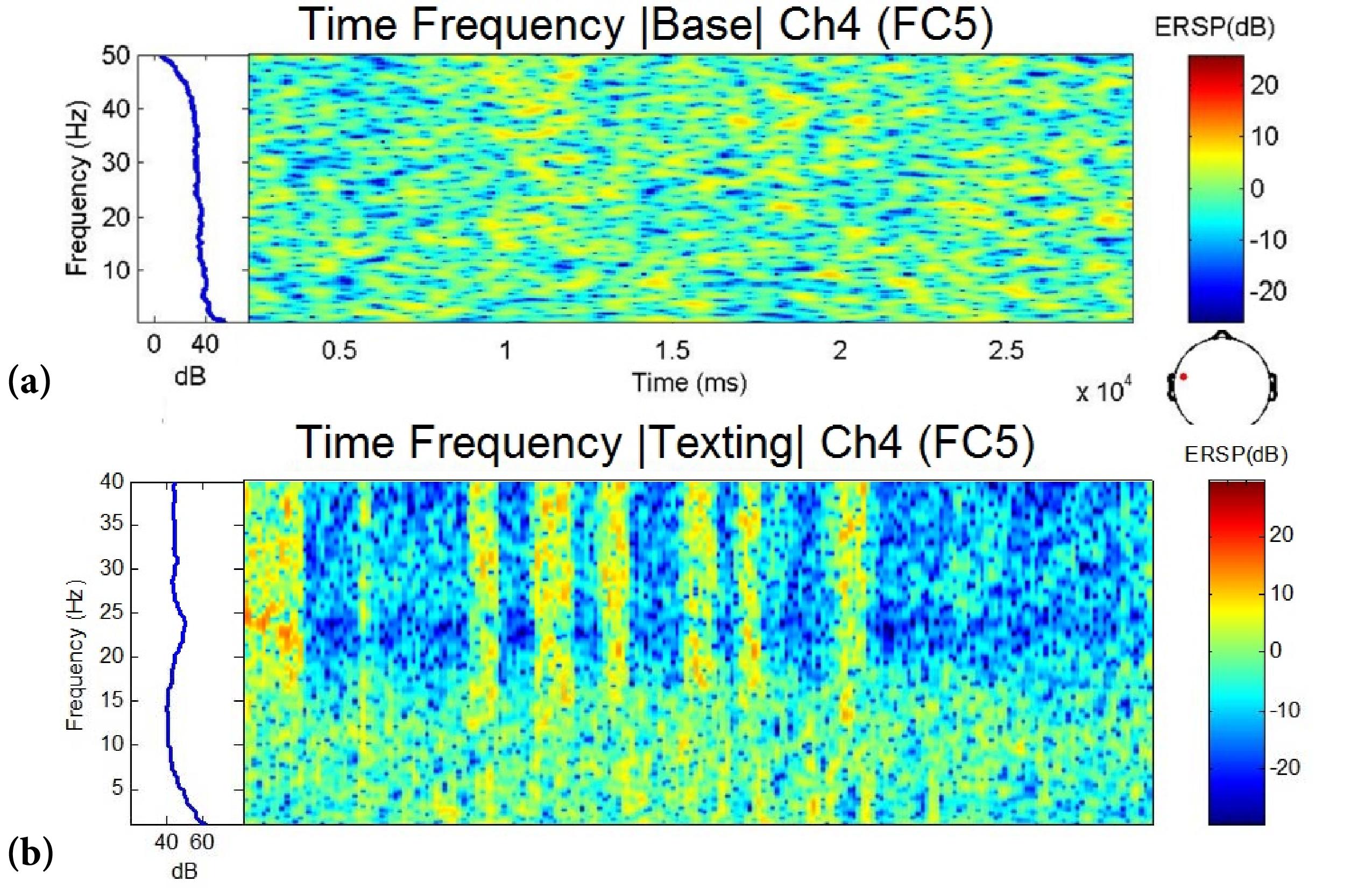}
  \caption{\textbf{Time-frequency plot of the FC5 scalp location comparing the baseline (a) and texting activity (b) of a subject while driving.} The left part of each figure shows power (dB) in the various EEG frequencies averaged over time, and the right part shows the detail time-frequency power spectrum. There was an overall increase in the total power of the texting activity. Also, the observed rhythmic appearance of high-low frequency in time domain EEG signals corresponded to a pattern mainly in the frequency range of 20 Hz to 40 Hz.}
  \label{fig:timefreq1}
\end{figure}

\begin{figure}[h!]
\centering
  \includegraphics[width=\linewidth]{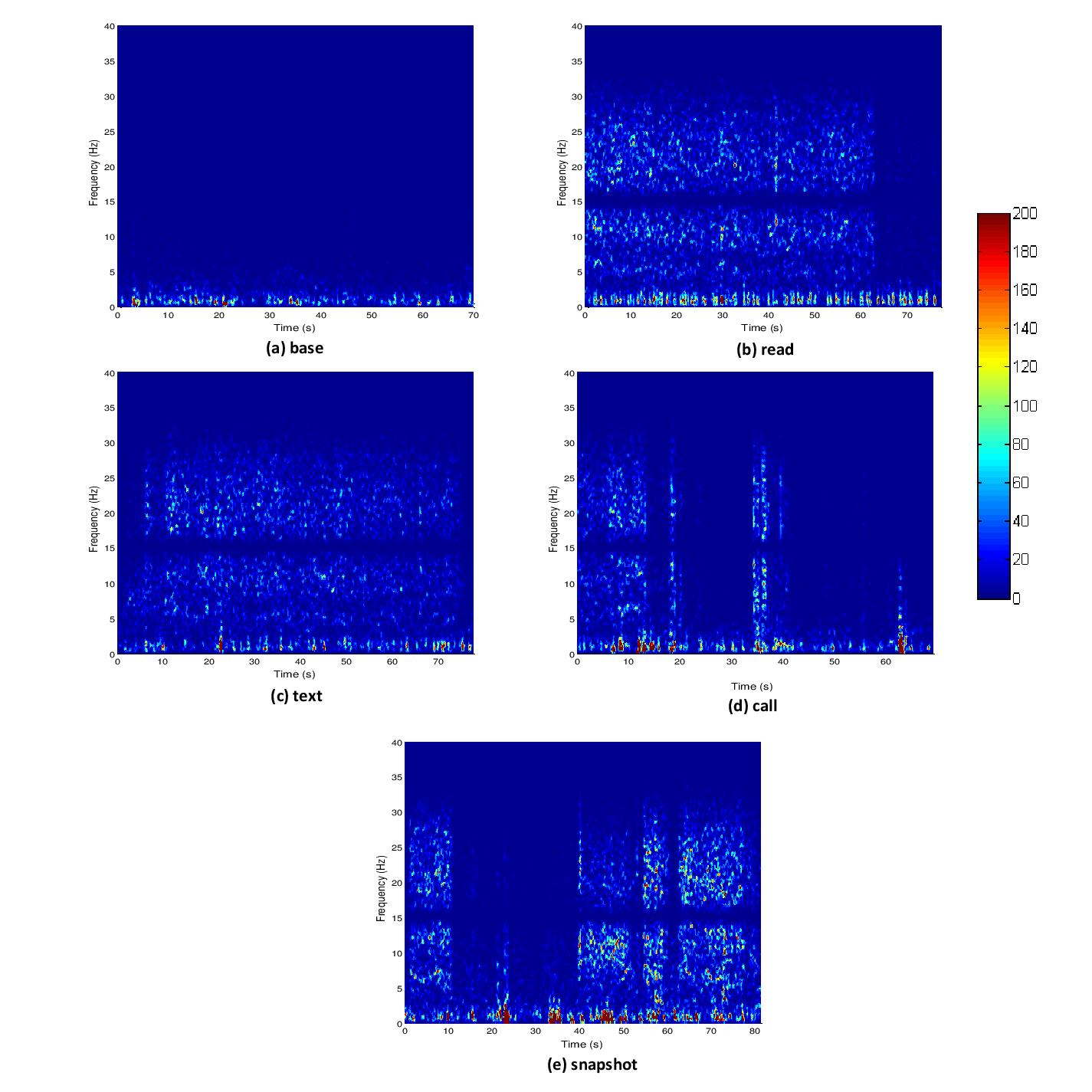}
  \caption{\textbf{Time-frequency representation of the distraction activities (read, text, call, snapshot) from one driving session of a subject recorded using the single electrode headband at FC5 location.} All the distracted driving maneuvers exhibited an increase in the total EEG power spectrum compared to baseline (undistracted driving).}
  \label{fig:timefreq3}
\end{figure}

Similar to the previous experiment with the 14 electrodes headset, EEG signal acquisition, and video recording were synchronized while collecting the data. The recording was started a few seconds before the distraction activity to verify the base profile. We observed a similar pattern of high and low alternating frequency bursts in the time-frequency data obtained from the single electrode band for other tasks such as reading while driving (Fig. \ref{fig:timefreq1}). 
Likewise, the time-frequency spectrum of other distraction tasks obtained from the single electrode also showed similar changes with different signatures in average power during the distraction time span (Fig. \ref{fig:timefreq3}). The change was spread across the higher frequency bands except the low frequencies between 1$-$4 Hz. EEG signals have inherent characteristics that lower frequency components have higher magnitude compared to higher frequency components. No difference observed in the intensity for frequencies above 10 Hz in baseline is because the magnitude of lower frequencies have masked the subtle changes in the higher frequency components. We observed this behavior in power spectrums of all the subjects. However, again for illustration purposes, the spectrum plots show activities from only a single subject. \emph{Thus, the power spectrum analysis further strengthened our hypothesis of identifying distraction from FC5 location (statistical tests detailed in Results and Discussion section).}
\subsection{\textbf{Independent Component Analysis}} \label{ICA}
\begin{figure}[h!]
\centering
  \includegraphics[width=0.7\linewidth]{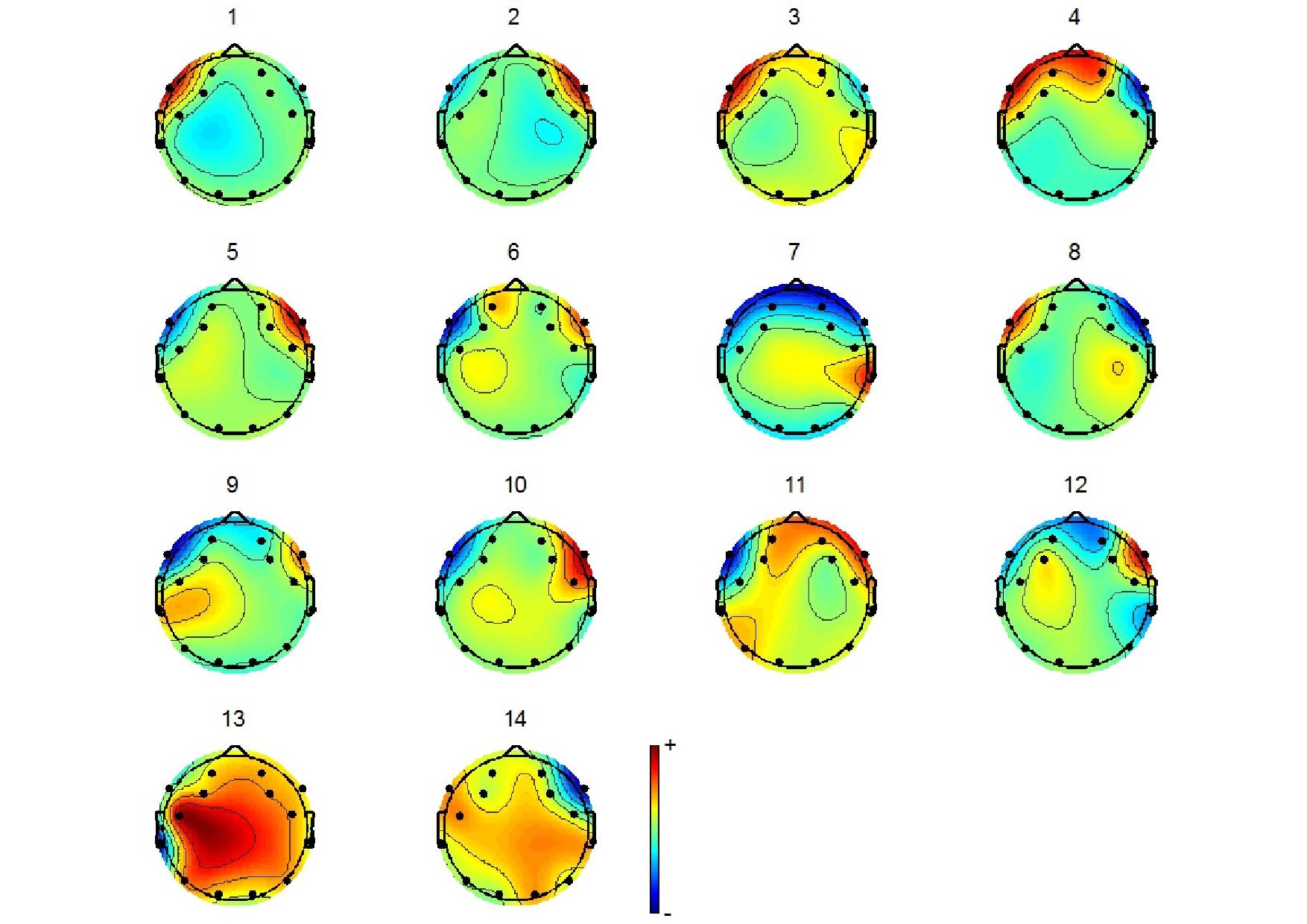}
  \caption{\textbf{2D scalp map projection for all the independent components.} These components were obtained from the independent component analysis (ICA) of the 14 channels EEG data for a texting trial of a subject. ICA helped to unmix the multi-channel EEG data into a sum of linearly independent, spatially fixed cortical sources.}
  \label{fig:ICA}
\end{figure}

\begin{figure}[h!]
\centering
  \includegraphics[width=0.7\linewidth]{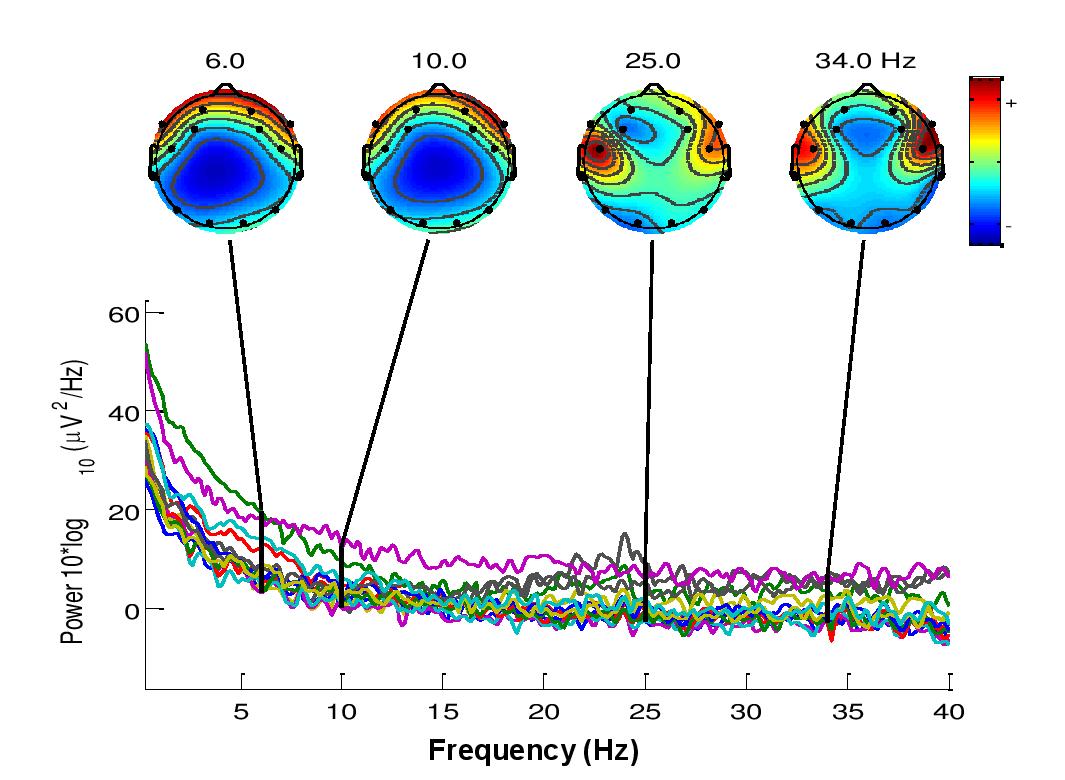}
  \caption{\textbf{The electrode spectra of the 14 channels (colored traces) and associated topographical distribution of power at the specified center frequencies of the EEG bands (theta-$\theta$, alpha-$\alpha$, beta-$\beta$, gamma-$\gamma$).}}
  \label{fig:cf}
\end{figure}

\begin{figure}[h!]
\centering
  \includegraphics[width=0.6\linewidth]{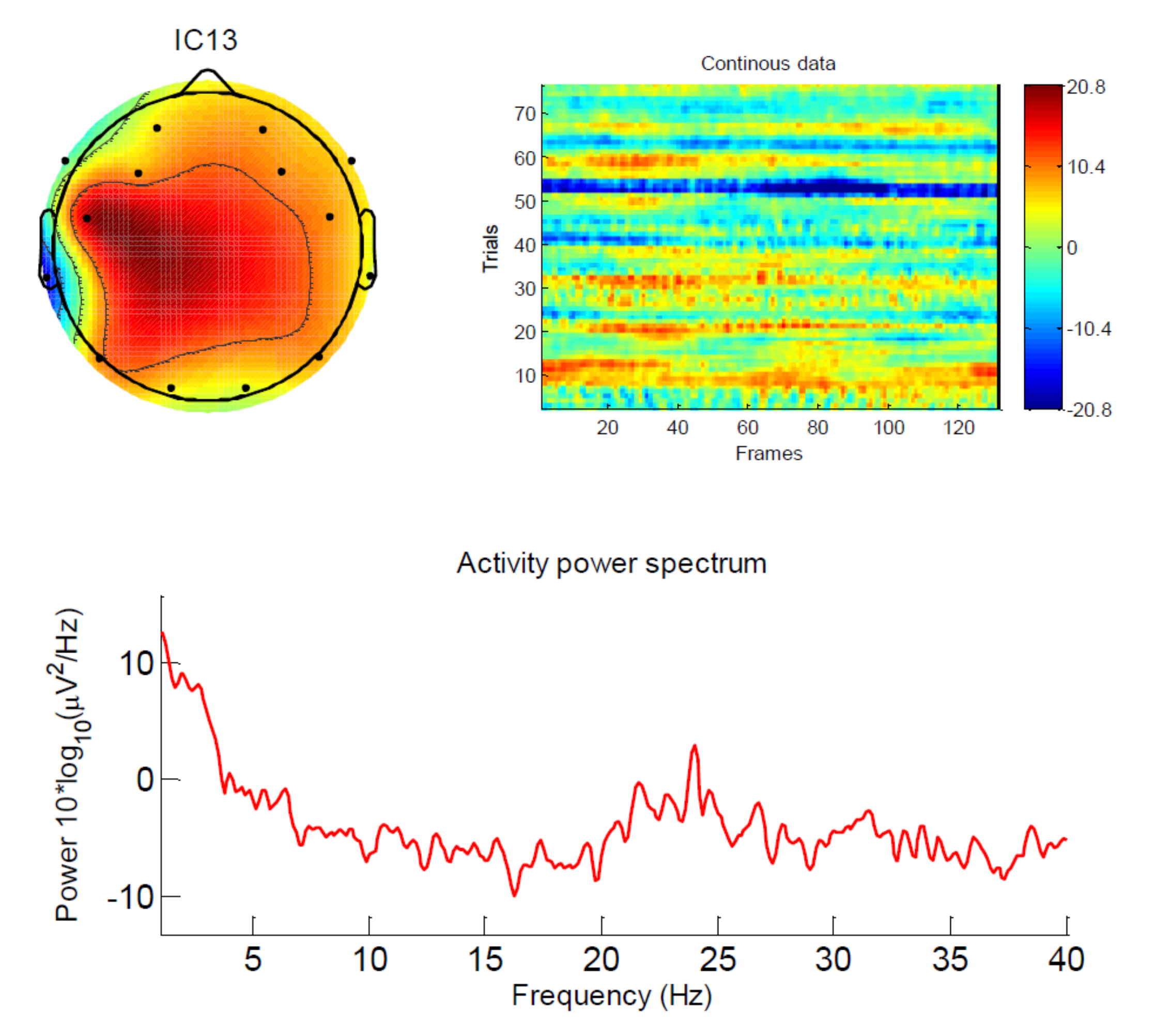}
  \caption{\textbf{Activity power spectrum and the scalp map projection of component 13 for a texting trial of a subject (same trial as in Fig. \ref{fig:ICA}).} This dipole-like scalp map distribution of component 13 has a beta band peak near 25 Hz and accounts for a significant EEG variance observed in the distraction activity of the texting trial.}
  \label{fig:comp1}
\end{figure}

\begin{figure}[h!]
\centering
  \includegraphics[width=0.6\linewidth]{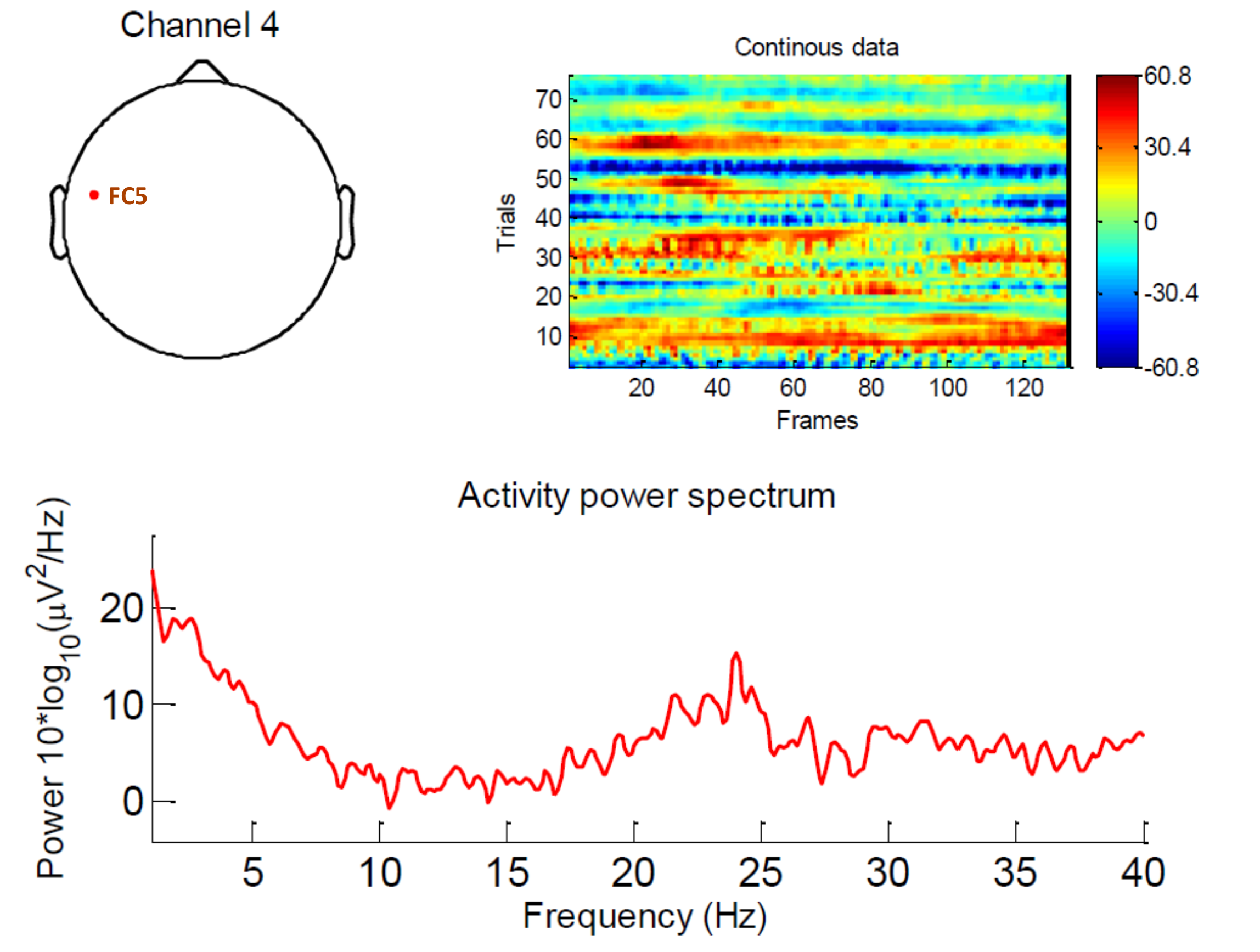}
  \caption{\textbf{Activity power spectrum of FC5 channel for the same texting trial of a subject as in Fig. \ref{fig:comp1}.} The power spectrum showed a perfect synchronization between the neural activations recorded at FC5 electrode and the cortex source identified by component 13 with a similar beta band peak near 25 Hz.}
  \label{fig:comp2}
\end{figure}

\begin{figure}[h!]
\centering
  \includegraphics[width=0.6\linewidth]{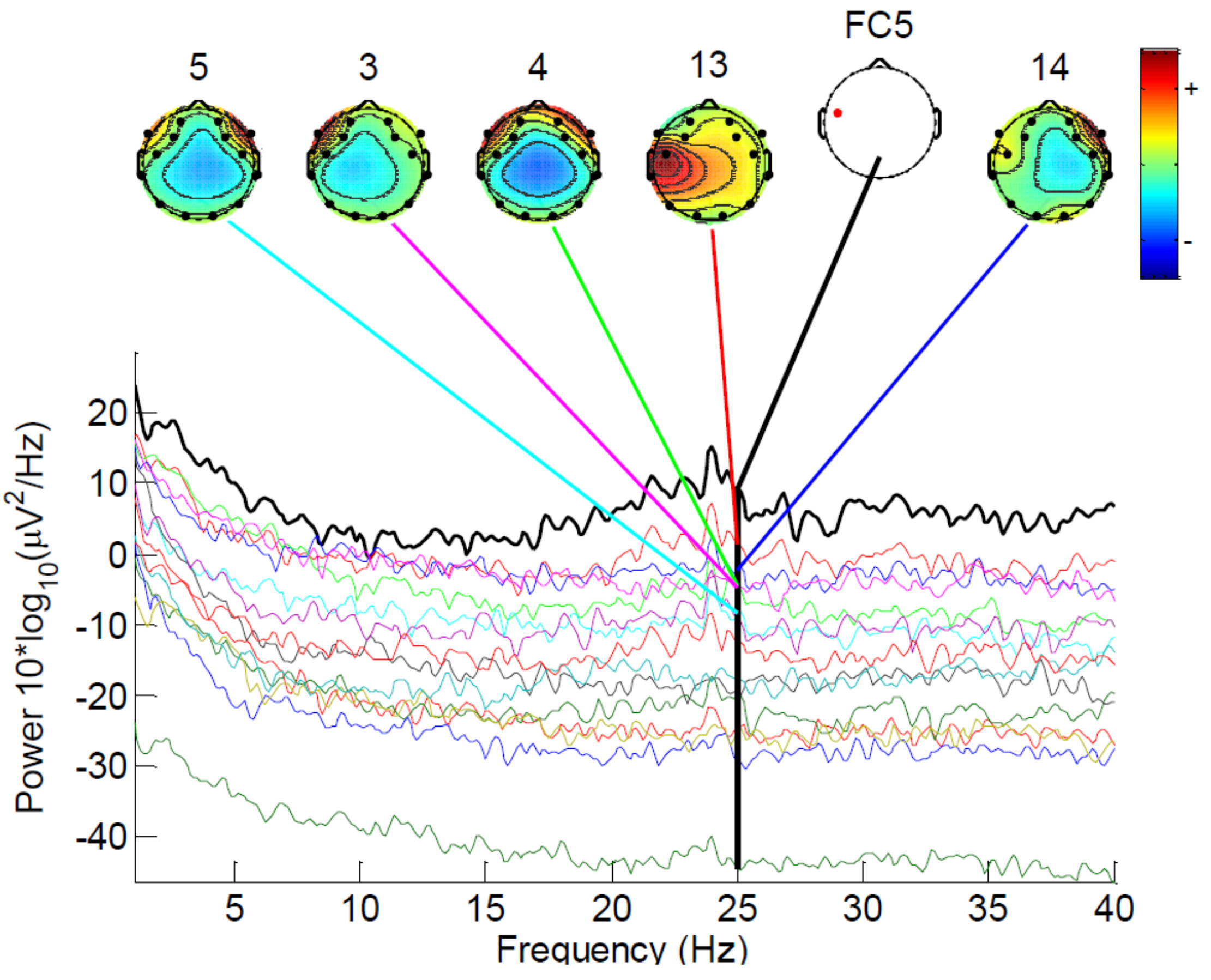}
  \caption{\textbf{Contributions of the five largest independent components as a percentage of the total power at FC5 electrode at frequency 25 Hz for the texting trial (same trial as Figure 12) of a subject.} Component 13 accounted for the maximum variance compared to other independent components with more than 87\% of the power at the FC5 electrode.}
  \label{fig:comp3}
\end{figure}

The 2D scalp map for each component of the independent component analysis (ICA) of all 14 channels is shown in Fig. \ref{fig:ICA}. Again, ICA was carried out for all the activities, but we discuss the components of one distraction type (texting) to facilitate our discussion. The components were calculated for the entire duration of a trial \cite{delorme2004eeglab}. ICA is a popular method to separate linearly mixed sources. However, even when the sources are not truly independent, it converges to a maximally independent space of sources separation, which was desired by our study. Component 13 accounts for a significant EEG variance observed in the distraction activity of texting. Component 4 contributes to the eye movement artifacts in the frontal region.
The ICA algorithm has no apriori knowledge about the electrode positions for the EEG signals and gives us maximally independent sources of cortical synchrony \cite{makeig1996independent, anemuller2003complex}. Fig. \ref{fig:cf} shows the topographical distribution of power at the center frequencies of the EEG bands (theta-$\theta$, alpha-$\alpha$, beta-$\beta$, gamma-$\gamma$) and each colored trace represents power spectrum of the texting activity in a channel. We can see the scalp distribution of power at 25 Hz (beta-$\beta$) and 34 Hz (gamma-$\gamma$) of a texting task were concentrated in regions around FC5 and FC6 electrodes.
Similarly, the scalp map distribution (Fig. \ref{fig:ICA}) of the components shows that IC 13 accounts for a significant variance in the activity spectrum. Also, it was evident that IC 13 has the highest contribution to the activity at FC5 electrode at 25 Hz as shown in Fig. \ref{fig:comp3}. We compared the activity spectra of all components at this electrode and contribution of IC 13 to other scalp locations to validate this (only two included here due to space constraints). The activity spectrum plots in Figs. \ref{fig:comp1} and \ref{fig:comp2} showed a perfect synchronization between the neural activations recorded at FC5 electrode and the cortex source identified by Component 13. Hence, it was reasonable to conclude that the activity observed in FC5 electrode could be used as a cortical source to identify distractions without any significant loss of distraction related information (ignoring the smaller contributions of other cortical sources to the power distributions in higher frequency bands such as beta and gamma). This favors the use of the single electrode (FC5) compared to that of bulky 14-electrode system.
\section{Stimuli-Response Quantification}
People develop different kinds of driving habits over time. The level of concentration, skill and reaction times vary among subjects. EEG signals of each person while driving are different (significant or minute), even if they are performing the same activities. Our goal was to study the effects of the various types of distraction stimuli on these signals. Therefore, we looked into the problem of classifying the distraction response for each subject separately through the existing machine learning models. Later, we shall discuss how this information can be used to calibrate an application to detect the distraction levels and create an index to measure such distractions.
\subsection{\textbf{Quantification}}
We studied the classification problem of EEG distraction response as a two-class problem and a five-class problem. In the two-class problem, the distraction class was the positive class \textendash a grouping of all the distraction tasks (read, text, call and snapshot). The other category, the base class was the negative class which represented a normal driving pattern of a subject. For the five-class problem, each of the tasks performed while driving were considered a separate category \textendash base, text, read, call, and snapshot and we performed multi-class machine learning assessment. Since each task recording was nearly 60 seconds, we split the baseline and distraction tasks to form multiple trials of three to five seconds. This sub-sampling of the tasks was done to increase the instances and have a balanced dataset for classification. Also, it was difficult to precisely perform a distraction activity for three to five seconds. Hence, the same distraction activity was carried out for a longer time and repeated in multiple sessions to reliably capture distraction signatures. Secondly, the variations in brain dynamics were better captured in smaller intervals than using the whole reading as a single event.

Additionally, we compare the performance of distraction classification between 14 channels and one channel (FC5) for the pilot study subjects. Table \ref{tab:data} shows the details of the complete dataset with the number of trials for each task. Subjects 1 and 2 performed the distraction maneuvers using 14 electrodes headset. Subject 15's data was recorded on the mobile phone to test the real-time application for distraction detection. Therefore, we label the data obtained from either the Brain-Computer-Interface (BCI) or the Brain-Mobile-Interface (BMI).

\subsection{\textbf{Feature Extraction}} \label{feature}
\begin{figure}[htb]
\centering
  \includegraphics[width=0.8\linewidth]{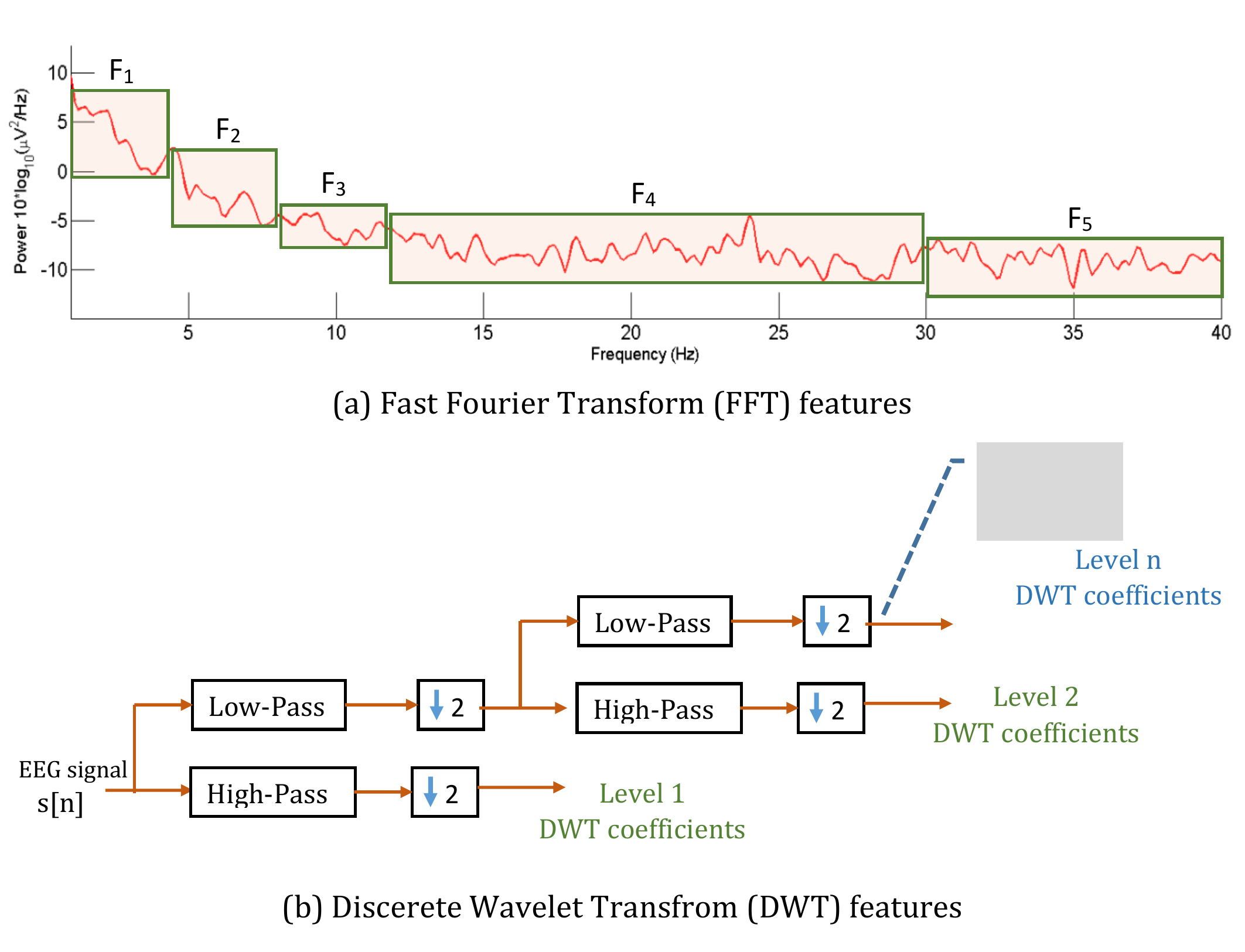}
  \caption{\textbf{The process of extracting features from the frequency spectrum of EEG to form feature vectors used in distraction classification.} Fast Fourier Transform (FFT) and Discrete Wavelet Transform (DWT) were the two methods used to obtain the relevant features. Using FFT (a), power values in each frequency band of the transformed signal were averaged for the feature vector. Using DWT (b), mean of the absolute values of the coefficients and average power of the coefficients in each level wavelet band were averaged to produce the feature vector.}
  \label{fig:fft}
\end{figure}

The feature extraction process was the same for the data obtained from both the 14 electrodes and single electrode headsets. We used two approaches for feature extraction.
\begin{itemize}
  \item \emph{Fast Fourier Transform (FFT):} First, the raw time domain data was converted to frequency-time spectrums. Then, the window size was increased to the whole EEG reading duration of a trial, so that it will eliminate the time axis of the plots and only produce the frequency spectrum. Thus, the time window for the FFT was the total EEG recording duration of a trial. The attributes consisted of 5 bands, namely Delta (1-4 Hz), Theta (4-8 Hz), Alpha (8-12 Hz), Beta (12-30 Hz) and Gamma (31-40 Hz). The power values in each window were averaged to produce the feature value.  Hence, for 14 channels there were $14\times5=70$ values in the final feature vector for each EEG reading. Similarly, for the single electrode we had five values in the feature vector of a trial (Fig. \ref{fig:fft}).
  \item \emph{Discrete Wavelet Transform (DWT):} We used the Daubechies family (db8) as the mother wavelet for the transform. It's irregular shape and compact nature help in analyzing signals with discontinuities and sharp changes such as EEG signals. Instead of using all the coefficients at each decomposition level, we extracted the information from the wavelet coefficients only at the levels corresponding to the five frequency bands mentioned in FFT feature extraction (Fig. \ref{fig:fft}). The mean of the absolute values of the coefficients and their average power in these levels were used as features. Thus, the DWT feature vector was composed of 10 features for an electrode.
\end{itemize}
Once the feature vectors were extracted from an each EEG reading, they were organized into an ARFF (Attribute-Relation File Format) file format for Weka \cite{weka} tool to process various machine learning models.
\begin{table*} [htb]
\captionsetup{width=\linewidth}
 \caption{Classification performance of the two class distraction problem for each subject measured with Bayesian Networks and Multilayer Perceptron using k-fold cross validation. The true class consisted of the distraction events, and the other class was the base driving (undistracted) activity.}
  \centering

    \begin{threeparttable}[b]

  \scalebox{0.7}{
    \begin{tabular}{c|*8c|} \hline \hline

     \ {Subject} & {Electrodes} & {Interface} & {Classifier} &  {Precision} & {Recall} & {Accuracy \%} & {F-Measure} & {AUC} \\ \hline
1	&	14	&	BCI	&	Bayesian	Network	&	0.962	&	0.958	&	95.83	&	0.958	&	0.986	\\	\hdashline	
1	&	one	&	BCI	&	Bayesian	Network	&	0.917	&	0.917	&	91.66	&	0.917	&	0.847	\\		
2	&	14	&	BCI	&	Multilayer	Perceptron	&	0.96	&	0.957	&	95.65	&	0.956	&	1	\\	\hline	
3	&	one	&	BCI	&	Bayesian	Network	&	0.955	&	0.95	&	95	&	0.95	&	0.97	\\		
4	&	one	&	BCI	&	Multilayer	Perceptron	&	0.951	&	0.947	&	94.73	&	0.946	&	1	\\		
5	&	one	&	BCI	&	Multilayer	Perceptron	&	1	&	1	&	100	&	1	&	1	\\		
6	&	one	&	BCI	&	Multilayer	Perceptron	&	0.907	&	0.889	&	88.88	&	0.886	&	0.775	\\		
7	&	one	&	BCI	&	Multilayer	Perceptron	&	0.929	&	0.92	&	92	&	0.918	&	0.923	\\		
8	&	one	&	BCI	&	Bayesian	Network	&	1	&	1	&	100	&	1	&	1	\\		
9	&	one	&	BCI	&	Multilayer	Perceptron	&	0.946	&	0.938	&	93.75	&	0.938	&	0.908	\\		
10	&	one	&	BCI	&	Bayesian	Network	&	0.904	&	0.882	&	88.23	&	0.88	&	0.757	\\		
11	&	one	&	BCI	&	Multilayer	Perceptron	&	0.944	&	0.938	&	93.75	&	0.937	&	0.898	\\		
12	&	one	&	BCI	&	Bayesian	Network	&	0.9	&	0.875	&	87.5	&	0.873	&	0.906	\\		
13	&	one	&	BCI	&	Bayesian	Network	&	0.889	&	0.84	&	84	&	0.843	&	0.774	\\		
14	&	one	&	BCI	&	Multilayer	Perceptron	&	0.864	&	0.813	&	81.25	&	0.806	&	0.695	\\		\hline
15	&	one	&	BMI	&	Multilayer	Perceptron	&	0.87	&	0.868	&	86.84	&	0.869	&	0.923	\\	\hline	\hline

    \end{tabular}
    }

     \begin{tablenotes}
    \item BCI - Brain-Computer-Interface, BMI - Brain-Mobile-Interface, AUC - Area under curve
  \end{tablenotes}
 \end{threeparttable}

    \label{tab:twoclass}
\end{table*}
\section{Results and Discussion}
\subsection{\textbf{Classification}}
The classification of EEG distraction response was performed using supervised learning models. We include the Brain-Computer-Interface (BCI) and Brain-Mobile-Interface (BMI) to compare the classification performance using both the interfaces. We used the standard probabilistic and neural network classifiers - Bayesian Network and Multilayer Perceptron as the machine learning methods because of their extensive use in EEG analysis \cite{anderson1996classification, palaniappan2005brain, lotte2007review, solhjoo2004mental}. K-fold cross validation was conducted for each classifier. The training set was randomly divided into K disjoint sets of equal size, K-1 folds used for training and the remaining one for testing with each time a different set held out as the test set. The overall performance was analyzed by using various indexes such as accuracy, precision, recall, F-score, and AUC (area under curve). F-score is a measure that combines precision and recall, and is computed as the harmonic mean of them.

\begin{figure}[h!]
\centering
  \includegraphics[width=0.85\linewidth]{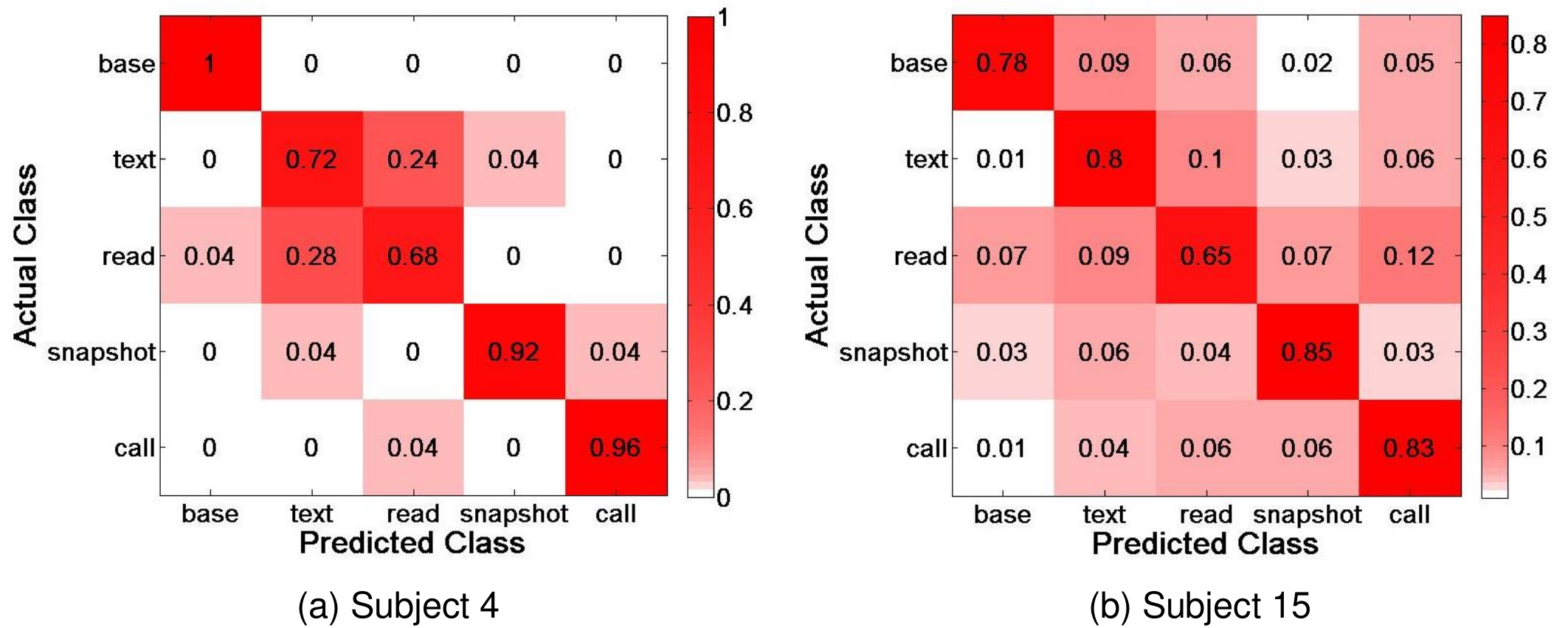}
  \caption{\textbf{Normalized confusion matrices of two subjects to depict the performance of the system to distinguish between various types of distraction (read, text, call, and snapshot).} The normalized counts on the diagonal are the true positives for each class and the counts not on the diagonal are the errors for each class. The classifier performance for a subject was independent of the other subjects.}
  \label{fig:cms}
\end{figure}

Tables \ref{tab:twoclass} and \ref{tab:fiveclass} show the distraction performance of the classifiers for the two-class (base, distract) and five-class (base, read, text, call, camera) problem respectively for each subject.
\begin{table*} [htb]
\captionsetup{width=\linewidth}
 \caption{Classification performance of the five class distraction problem for each subject measured with Bayesian Networks and Multilayer Perceptron using k-fold cross validation. Each of the tasks performed while driving were considered a separate class - base (undistracted), text, read, call, and snapshot.}
  \centering
  \begin{threeparttable}[b]
  \scalebox{0.7}{
    \begin{tabular}{c|*8c|} \hline \hline
       \ {Subject} & {Electrodes} & {Interface} & {Classifier} &  {Precision} & {Recall} & {Accuracy \%} & {F-Measure} & {AUC} \\ \hline
1	&	14	&	BCI	&	Bayesian	Network	&	0.961	&	0.947	&	94.73	&	0.946	&	1	\\	\hdashline		
1	&	one	&	BCI	&	Bayesian	Network	&	0.876	&	0.871	&	87.09	&	0.871	&	0.993	\\			
2	&	14	&	BCI	&	Multilayer	Perceptron	&	0.941	&	0.929	&	92.85	&	0.929	&	0.981	\\	\hline		
3	&	one	&	BCI	&	Multilayer	Perceptron	&	0.857	&	0.848	&	84.8	&	0.847	&	0.956	\\			
4	&	one	&	BCI	&	Multilayer	Perceptron	&	0.856	&	0.856	&	85.6	&	0.856	&	0.966	\\			
5	&	one	&	BCI	&	Bayesian	Network	&	0.804	&	0.8	&	80	&	0.803	&	0.953	\\			
6	&	one	&	BCI	&	Multilayer	Perceptron	&	0.728	&	0.728	&	72.8	&	0.727	&	0.921	\\			
7	&	one	&	BCI	&	Multilayer	Perceptron	&	0.816	&	0.813	&	81.33	&	0.814	&	0.95	\\			
8	&	one	&	BCI	&	Multilayer	Perceptron	&	0.724	&	0.7	&	70	&	0.698	&	0.881	\\			
9	&	one	&	BCI	&	Multilayer	Perceptron	&	0.835	&	0.768	&	76.8	&	0.777	&	0.92	\\			
10	&	one	&	BCI	&	Bayesian	Network	&	0.823	&	0.792	&	79.2	&	0.794	&	0.935	\\			
11	&	one	&	BCI	&	Multilayer	Perceptron	&	0.726	&	0.724	&	72.4	&	0.719	&	0.882	\\			
12	&	one	&	BCI	&	Multilayer	Perceptron	&	0.71	&	0.672	&	67.2	&	0.676	&	0.873	\\			
13	&	one	&	BCI	&	Multilayer	Perceptron	&	0.607	&	0.6	&	60	&	0.599	&	0.837	\\			
14	&	one	&	BCI	&	Multilayer	Perceptron	&	0.665	&	0.651	&	66	&	0.667	&	0.868	\\ \hline			
15	&	one	&	BMI	&	Multilayer	Perceptron	&	0.79	&	0.789	&	78.86	&	0.789	&	0.918	\\	\hline	\hline	
    \end{tabular}
    }
     \begin{tablenotes}
    \item BCI - Brain-Computer-Interface, BMI - Brain-Mobile-Interface, AUC - Area under curve
  \end{tablenotes}
 \end{threeparttable}
    \label{tab:fiveclass}
\end{table*}
Similarly, we analyzed the complete dataset consisting of all the subjects together to study the two-class and the five-class problem (Table \ref{tab:allclass}). The performance measures were weak compared to the subjective classification of distraction because of the difference in the EEG response among the subjects while performing similar tasks.
\begin{table*} [htb]
\captionsetup{width=\linewidth}
 \caption{Overall classification performance of the two class and the five class distraction problem using the trials from all subjects. Baseline (undistracted) driving, and collectively the distracted driving activities constituted the two class problem. The various tasks performed while driving - base (undistracted), text, read, call, and snapshot separately formed the five class problem.}
  \centering
 \scalebox{0.8}{
    \begin{tabular}{c|*6c|} \hline \hline
Classifier	& 	Classes	& 	Precision	& 	Recall	& 	Accuracy (\%)	& 	F-measure	& 	AUC	\\	\hline
Bayesian Network	& 	Two class	& 0.776   & 0.774	& 	77.37	& 	0.774	& 	0.81	\\	
	& 	Five class	& 	 0.502   &  0.509 	& 	50.93	& 	0.498	& 	0.816	\\	\hline
Multilayer Perceptron	& 	Two class	& 0.81 & 0.81   & 	80.99	& 	0.809	& 	0.803	\\	
	& 	Five class	&  0.511 &  0.511 &  51.06 & 0.51  &    0.792 \\	\hline
 \end{tabular}
    }
    \label{tab:allclass}
\end{table*}
Five-class classification measures were worse than the two-class problem. The normalized confusion matrices of the five-class problem were compared among different subjects (Fig. \ref{fig:cms} shows for two subjects). It clearly reflects the variation in the classification of the same distraction activities for the subjects. Hence, a real-time distraction detection application requires calibration for each driver, as discussed in the section below.

\subsection{\textbf{Distraction Index}} \label{index}
We can observe the differences in the active regions of the frequency bands for various distraction activities. However, it's difficult to quantify these distractions from raw EEG signals, time-frequency plots or even machine learning models, especially for real-time alerts. Therefore, we developed an index of distraction using the signal power in various frequency bands of EEG as follows;
 \[ Distraction Index (DI)  = \theta / \alpha + \alpha / \beta + \beta/ \gamma \]
where $\theta$ is the average EEG power between 4-8 Hz, $\alpha$ - between 8-12 Hz, $\beta$ - between 12-30 Hz, and $\gamma$ - between 31-40 Hz.

\begin{figure}[htb]
\centering
  \includegraphics[width=0.95\linewidth]{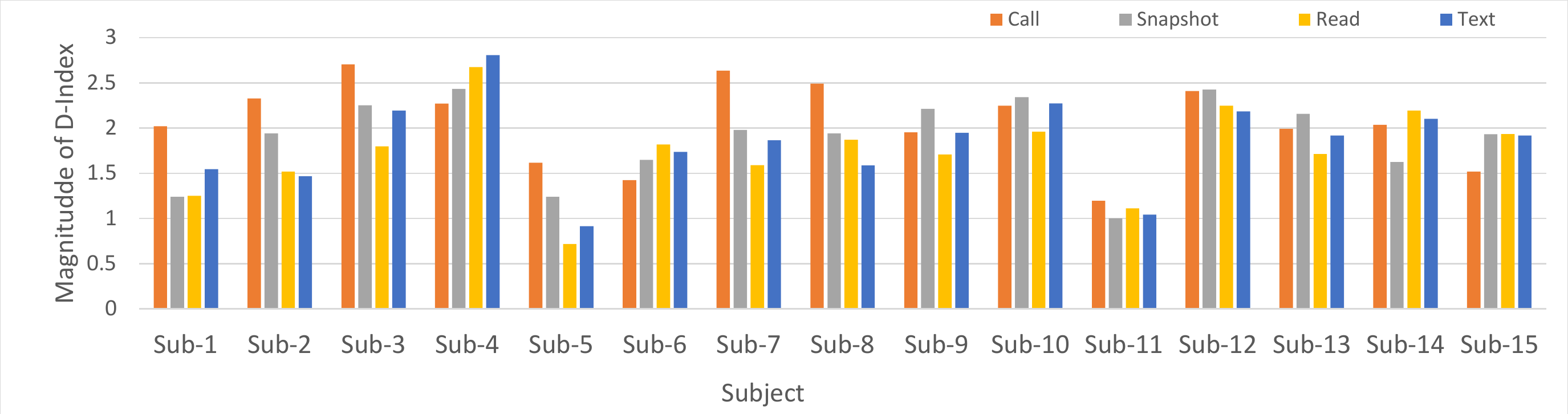}
  \caption{\textbf{Distraction Index (D-Index) of various types of distractions (read, text, call, and snapshot) for each subject.} Features from the time-frequency spectrum were used to rank these distractions. The rank was obtained from the value of distraction index given by a summation of power ratios obtained in different frequency spectrums of the EEG for an activity ($\theta / \alpha + \alpha / \beta + \beta/ \gamma$).}
  \label{fig:ind}
\end{figure}

The reasons to include these EEG frequency bands was to capture their varying contribution in distraction. The ratios were important because the common noise in all the bands such as muscle artifacts will be mitigated. EEG signals have inherent characteristics that lower frequency components have higher magnitude compared to higher frequency components. Hence, each term (e.g. $\theta$/$\alpha$) in the index was specifically a ratio of the adjacent band so that no band masks the contribution of another frequency band. The addition of the three ratios increases the dynamic range of the index and helps in evaluating the extent of distraction. We also ranked the tasks in order of the severity of distraction and compared it to the actual order of distractions rated by the subjects (Table \ref{tab:quest} and Fig. \ref{fig:ind}). An average over all the trials was taken to represent the mean Distraction Index for each task of a subject. The differences in the levels of distraction were certainly unique to the activity and subject.

\begin{table*} [htb]
\captionsetup{width=\linewidth}
 \caption{Users' perception of the level of distraction for the various driving tasks compared to the order obtained by our distraction index metric (Figure 18).}
  \centering
  \scalebox{0.6}{
    \begin{tabular}{c|c|c|c|c|c} \hline \hline

Subject	& 	Tasks in descending  & Least distraction & Subject	& 	Tasks in descending   & Least distraction \\
		&	order of distraction         &  task 		   &	        &   order of distraction  & task  \\ \hline
1	& 	Snapshot, Text, Read, Call	& 	None	& 9	& 	Snapshot, Text, Read, Call	& 	Call	\\		
2	& 	Read, Text, Snapshot, Call	& 	None	& 10	& 	Text, Snapshot, Read, Call	& 	Call	\\			
3	& 	Snapshot, Text, Read, Call	& 	Call	& 11	& 	Snapshot, Read, Text, Call	& 	Call	\\			
4	& 	Text, Snapshot, Read,  Call	& 	None	& 12	& 	Snapshot, Read, Text, Call	& 	Call	\\			
5	& 	Text and Read, Snapshot, Call 	& 	Call	& 13	& 	Snapshot, Text, Read, Call	& 	None	\\				
6	& 	Text, Read, Snapshot, Call	& 	None	& 14	& 	Read, Snapshot, Text, Call	& 	None	\\			
7	& 	Read, Text, Snapshot, Call	& 	Call	& 15	& 	Snapshot, Read, Text, Call	& 	None	\\
8	& 	Read, Text, Snapshot,  Call	& 	None	&       &  	                            &           \\	\hline \hline

 \end{tabular}
   }
    \label{tab:quest}
\end{table*}
\subsection{\textbf{Statistical Analysis}}
Nonparametric test - Wilcoxon matched pairs signed rank test \cite{marusteri2010comparing} was used to compare between the base and distraction activities as we had two paired groups with sample size 15.
\textbf{H\_0 (null hypothesis):} There was no difference in the average EEG power  of the subjects in normal (base) driving compared to distracted driving scenario. The critical value for this two-sided test with n=15 (sample size) and $\alpha$=0.05 (level of significance) is 25 and the decision rule is to reject H\_0 if $W \le 25$ (test statistic) \cite{wilcox}. We obtained p = 6.1035e-05 and signed rank (W) = 0. Since $W \le 25$, we can reject the null hypothesis that there was no difference between the average power spectrums of the two groups.
Table \ref{tab:stat1} shows the average EEG power in the activity spectrum of 15 subjects during normal/base driving and various distracted driving scenarios.

\begin{table*} [htb]
\captionsetup{width=\linewidth}
 \caption{Average EEG power ($\mu V^2/Hz$) in the activity spectrum of 15 subjects during normal/base driving and various distracted driving scenarios.}
  \centering
  \scalebox{0.8}{
    \begin{tabular}{c|c|c|c|c|c} \hline \hline

Subject	&	Baseline	&	Distraction & Subject	&	Baseline	&	Distraction	\\ \hline
1	&	4.322328629	&	8.106168937	 &    9	        &	4.449198532	&	5.391230252	\\
2	&	2.664582	&	3.4866521	 &    10	    &	2.604087967	&	6.304723748	\\
3	&	3.111189679	&	7.875135417	 &    11    	&	1.803099442	&	3.435682585	\\
4	&	2.086580297	&	9.029101508	 &    12	    &	2.150271466	&	8.187098718	\\
5	&	2.336924079	&	5.267485052	 &    13	    &	3.682609891	&	4.267558023	\\
6	&	1.585887914	&	4.56627038	 &    14	    &	3.857001751	&	4.310074392	\\
7	&	1.836878292	&	2.757522033	 &    15	    &	3.346612251	&	6.025901425	\\
8	&	1.083730232	&	2.590520998	 &              &               &               \\ \hline \hline

 \end{tabular}
   }
    \label{tab:stat1}
\end{table*}

\begin{table*} [htb]
\captionsetup{width=\linewidth}
 \caption{Average EEG power ($\mu V^2/Hz$) measured at various electrodes (14) for a subject performing different distraction tasks while driving.}
  \centering
 \scalebox{0.6}{
    \begin{tabular}{c|*8c|} \hline \hline
Activity  	&	Trial	&	AF3	&	F7	&	F3	&	FC5	&	T7	&	P7	&	O1	\\ \hline
	&	1	&	2.4214752	&	2.4981809	&	2.3926721	&	2.4672167	&	2.4537251	&	2.5601602	&	2.8574317	\\
Call	&	2	&	2.4347413	&	2.4729524	&	2.4361315	&	2.3754349	&	2.4241006	&	2.6834061	&	2.7273037	\\
	&	3	&	2.4279513	&	2.4650245	&	2.4330645	&	2.4140978	&	2.4705017	&	2.522824	&	2.676507	\\
	&	4	&	2.5287697	&	2.5791395	&	2.5299034	&	2.517122	&	2.4597104	&	2.8035517	&	3.1059065	\\
	&	1	&	2.6744332	&	4.0664635	&	2.6502247	&	4.4818392	&	3.8846357	&	2.7514007	&	2.7116179	\\
Read	&	2	&	2.9250038	&	2.9563489	&	2.8790686	&	3.3520169	&	3.1556339	&	2.7855651	&	2.7135143	\\
	&	3	&	2.635082	&	3.5064738	&	2.5460846	&	5.0874629	&	3.5924051	&	2.6608219	&	2.599179	\\
	&	4	&	2.6065793	&	3.4197419	&	2.6382325	&	3.9598022	&	2.9867475	&	2.5310094	&	2.4755123	\\
	&	1	&	3.2478967	&	3.5448542	&	3.9311559	&	4.0561976	&	3.4324446	&	3.5578644	&	3.6534033	\\
Snapshot	&	2	&	2.8223073	&	4.2352448	&	2.8834374	&	4.7531309	&	2.7212167	&	2.6228042	&	3.3407328	\\
	&	3	&	2.6893418	&	2.9432366	&	2.7538092	&	3.6618721	&	2.5144477	&	2.415884	&	2.4211514	\\
	&	4	&	2.6671553	&	3.0878718	&	2.7727659	&	3.7537279	&	2.5740485	&	2.4142389	&	2.3955367	\\
	&	1	&	2.9869101	&	4.2868524	&	3.1079965	&	5.8076501	&	3.3425231	&	2.6736579	&	2.6249413	\\
Text	&	2	&	2.5955961	&	2.7717819	&	2.4941945	&	2.9578588	&	2.5736146	&	2.3959432	&	2.413789	\\
	&	3	&	2.4637313	&	3.039103	&	2.547775	&	3.0722754	&	2.5503836	&	2.4926076	&	2.3438687	\\
	&	4	&	2.4745536	&	2.4191566	&	2.4521291	&	2.3884513	&	2.4261627	&	2.430717	&	2.3713632	\\ \hline
Activity  	&	Trial	&	O2	&	P8	&	T8	&	FC6	&	F4	&	F8	&	AF4	\\ \hline
	&	1	&	2.8621693	&	2.5580797	&	2.3760216	&	2.4576824	&	2.4028533	&	2.4821734	&	2.4983783	\\
Call	&	2	&	2.7133338	&	2.5219572	&	2.4690683	&	2.4598217	&	2.4115481	&	2.4615436	&	2.4418302	\\
	&	3	&	2.6805191	&	2.5085919	&	2.4453657	&	2.3933685	&	2.4249372	&	2.4244306	&	2.4233027	\\
	&	4	&	3.0501297	&	2.8355846	&	2.6063755	&	2.5199046	&	2.4523113	&	2.494493	&	2.5130088	\\
	&	1	&	2.7799065	&	2.7040005	&	2.6929064	&	3.6230438	&	2.6564288	&	5.0124846	&	2.9661417	\\
Read	&	2	&	2.8175149	&	2.426374	&	2.5852654	&	2.8593514	&	2.6942704	&	3.2400942	&	2.9929171	\\
	&	3	&	2.614526	&	2.5878289	&	2.6682968	&	3.1144288	&	2.6079624	&	3.3645532	&	2.7210817	\\
	&	4	&	2.5711627	&	2.4911165	&	2.5489254	&	3.1336505	&	2.518424	&	4.1870437	&	2.6842604	\\
	&	1	&	3.839448	&	3.120744	&	3.3499269	&	3.3758676	&	3.3225701	&	3.4972935	&	3.2727661	\\
Snapshot	&	2	&	3.1243985	&	2.7641568	&	2.6531153	&	2.9631443	&	2.6733487	&	3.0371733	&	2.7402406	\\
	&	3	&	2.4218996	&	2.4387803	&	2.4160261	&	2.6179724	&	2.6229534	&	2.7957058	&	3.064657	\\
	&	4	&	2.4215443	&	2.4266396	&	2.4192357	&	2.5572968	&	2.5817022	&	2.8330786	&	2.8972125	\\
	&	1	&	2.6944857	&	2.6453888	&	2.9356737	&	4.284204	&	3.0896397	&	5.5555711	&	3.5289237	\\
Text	&	2	&	2.4593246	&	2.386148	&	2.4703412	&	2.6398413	&	2.4761744	&	3.1388083	&	2.6252344	\\
	&	3	&	2.4223473	&	2.4383831	&	2.5079806	&	2.8174579	&	2.4746096	&	3.0528491	&	2.7614758	\\
	&	4	&	2.411974	&	2.4470479	&	2.3850336	&	2.4338882	&	2.4373732	&	2.373611	&	2.5013528	\\ \hline \hline

 \end{tabular}
    }
    \label{tab:stat2}
\end{table*}

Nonparametric Friedman's test \cite{marusteri2010comparing} was used to examine whether performing distraction tasks while driving caused any difference in the average EEG power measured at various electrodes (14) for a subject using a two-way layout as shown in Table \ref{tab:stat2}.

\textbf{H\_0 (null hypothesis):} There was no difference in the average EEG power measured at the 14 electrodes for the various distraction distractions performed by a subject with $\alpha$=0.05 (level of significance). We obtained that there was a statistically significant difference in the average EEG power of various electrodes during distraction while driving, $\chi^2(2)$ = 34.54, p = 0.001.

To examine where the differences actually occurred, we ran separate post hoc test - Wilcoxon signed-rank tests on the three combinations based on our previous observations. We used one-tailed test with n=16 (sample size) and $\alpha$ = (0.05/3 = 0.0166) (Bonferroni corrected level of significance). The summary of the test is as follows:
\begin{itemize}
\item \textbf{FC5 to FC6} - Test revealed an increase in the average EEG power at FC5 compared to FC6 (p = 0.0015, zval: 2.9733)
\item \textbf{FC5 to O1} - Test revealed an increase in the average EEG power at FC5 compared to O1 (p = 0.0039, zval: 2.6630)
\item \textbf{FC5 to O2} - Test revealed an increase in the average EEG power at FC5 compared to O2 (p = 0.0070, zval: 2.4562)
\end{itemize}

\subsection{\textbf{Discussion}}
Previous studies have revealed the effects of distracted driving regarding the attentional resources of one brain region being compromised over another. The observations indicated a substantial shift in brain activations from posterior (back of head area) to anterior (forehead area) regions, particularly in prefrontal area that is critical to driving \cite{wickens2008multiple} \cite{dux2006isolation} \cite{kramer2007neuroergonomics}. In our study through EEG signals, we also observed similar demands on the mental processing of different tasks while driving.

\textbf{\emph{Qualitative analysis}} of the subjects' EEG signals using time domain, frequency domain and independent component analysis led to FC5 scalp location as the most appropriate identifiable location for determining a distraction event (Figs. \ref{fig:video}, \ref{fig:timefreq1}, and \ref{fig:comp3}). The machine learning results were in agreement with our observations from the distraction analysis. The mean classification results using one electrode for both the two-class (91.54 $\pm$ 5.23)\% and five-class problems (76.99 $\pm$ 8.63)\% were reasonable across all the subjects (Tables \ref{tab:twoclass} and \ref{tab:fiveclass}). The F-measures and ROC areas were reasonably high as well (greater than 0.7). We see that the classification accuracy dropped from 95.83\% to 91.66\% using only the single electrode for the two-class problem in the pilot study for subject 1. However, the processing overhead and overall training time was significantly reduced from 14 electrodes down to one electrode with a small compromise in classification results. On the other hand, distinguishing between different kinds of distractions using a single electrode was relatively difficult as reflected by the accuracies of the five-class problem for all the subjects (Table \ref{tab:fiveclass}). There was a significant drop in classifier performance for subject one, from 94.73\% (14 electrodes) to 87.09\% (one electrode). The normalised confusion matrix of classification highlights these variations in detecting different types of distractions (Figure \ref{fig:cms}). For instance, the subtle differences in brain activations during a read and text activity while driving for subject 4 were hard to capture by the feature extraction mechanisms employing a single electrode. Nearly 25\% of the time classifier misclassified text as a reading activity and vice-versa.

The \textbf{\emph{individual differences}} in classification results have interesting implications. It suggests that the brain activations of each subject are different (significant or minute) even when they are carrying out the same distraction activity. The machine learning outcomes of training the classifiers with data from all the subjects help us to understand this (Table \ref{tab:allclass}). We clearly observed that there was no improvement in mean classification results of the two-class problem. The performance of the classifiers was worse even with the increase in the training and testing instances for the five-class problem (read, text, call and snapshot). It clearly signifies that a general classifier cannot be obtained for the entire population to provide metrics for distracted driving by just increasing the amount of training data. Hence, the need for an application that provides individual calibration was tested using the data from subject 15, acquired by the brain-mobile-interface (BMI). We achieved 86.84\% and 78.86\% accuracy for the two-class and the five-class problems respectively. Wang et al. \cite{wang2015online} showed the potential to predict the start of a map viewing distraction event during navigation while driving. Their study produced interesting results with nearly 81\% accuracy of prediction, however, conducted only in a simulated environment with 36 active EEG electrodes and a single distraction task. A combination of our mobile phone based detection application coupled with their prediction algorithm could form a robust framework for predicting driver intentions a-priori to the distraction events.

The \textbf{\emph{feature vectors}} also play a significant role in the overall classification of the distraction events. We observed that the DWT features alone gave good classification accuracies for the two-class problem. So, the size of feature vector was reduced by deselecting the FFT features. However, we needed more features for the five-class problem to improve the classification. Hence, the combined FFT-DWT feature vector was used for the classification of various distraction tasks, although, using both DWT and FFT feature vectors will add overhead in the application for real-time alerts. In addition to the features selection, standard probabilistic and neural network classifiers such as Bayesian Network and Multilayer Perceptron were chosen as the machine learning methods because of their application to diverse BCI problems such as motor imagery and mental tasks \cite{anderson1996classification, palaniappan2005brain, lotte2007review, solhjoo2004mental}. Through our analysis, we also observed that these two classifiers performed better in classifying distraction and it's types. The overall performance of these classifiers varied among the subjects. Comparing all the measures - recall, precision, accuracy, F-measure and ROC area, we observed that Multilayer Perceptron achieved better classification results for many subjects in our dataset, even though Bayesian Networks offered a computationally efficient approach. Therefore, classifier selection also showed a dependence on the individual differences in the synchronous brain activations of the drivers'.

\textbf{\emph{Our findings suggest that quantifying drivers' distraction responses and generating real-time alerts is possible through a single electrode placed at FC5 location.}} The system offers a comfortable, wearable EEG option without compromising mobility, and usability. A limitation of this study is the assumption that the tasks while driving (read, text, call and snapshot) were the only source of distraction in a given frame of time. Our replication of the on-road driving incorporates many other complexities that are not addressed by simulated driving such as anxiety and attention that are absent in a virtual environment as there are no real consequences of distraction tasks. Another limitation is that we did not test the cognitive abilities of the subjects before the data collection. The subjects confirmed that they had no known neurological disorders before participating in the study. Also, our current study did not involve younger (\textless 24 yrs) and older drivers (\textgreater 55 yrs), that may be an interesting comparison for further study.

The Brain Mobile Application Interface describes the prototype of the driver alert system that can communicate with the EEG headset, retrieve data, and perform the necessary signal processing to generate safe feedbacks to the distracted driver in real-time. In the following section we discuss the prototype of the driver alert system to achieve this goal.

\section{Brain Mobile Application Interface} \label{BMAPI}
We intended to use mobile computing resources to extract useful cognitive information from the brain signals to determine the real-time distraction level of a driver. Therefore, we developed a reliable interface between the brain sensor and the mobile phone. It provides ease of mobility for recording and processing the data in any driving environment. An Android API was used to display the raw EEG data and power in various frequency bands simultaneously while the data was being collected. We performed onboard processing in the phone to analyze, and compare these power levels in various EEG frequency bands. The expected primary results were displayed using the existing API; otherwise the rest of the data was stored in the phone memory or sent to a cloud \cite{5}. The cloud analysis of this mobile data can be advantageous to not only to the current driver but also the other vehicles/drivers for vehicle to vehicle (V2V) communication network in future.

\subsection{\textbf{Design}}
The brainwaves were captured via a single EEG sensor \cite{Neurosky}. The formula for converting raw EEG values to voltage is given by equation below \cite{10};
 \begin{equation}\label{1}
  Scalp Voltage =  \frac{{Rawvalue}\times \frac {(input voltage)}{4096}} {2000 }
 \end{equation}
where the input voltage is 1.8v, 2000x is the gain, and 4096 is the value range.

\begin{figure}[ht]
\centering
  \includegraphics[width=0.5\linewidth]{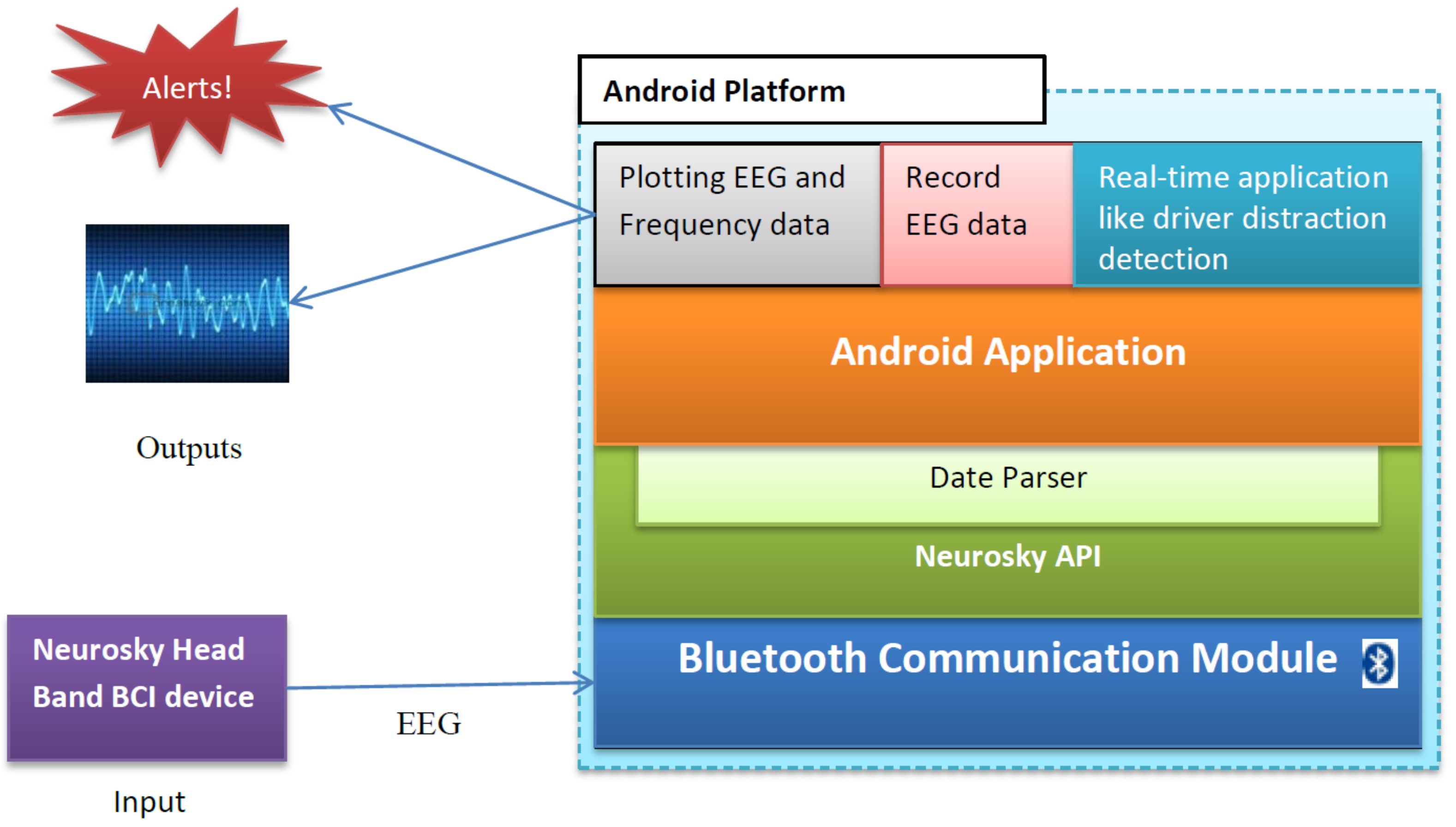}
  \caption{\textbf{The architecture of the Brain Mobile system used to obtain and process EEG data on the smartphone using an Android application interface.}}
  \label{fig:arch}
\end{figure}

\begin{figure}[htb]
\centering
  \includegraphics[width=0.6\linewidth]{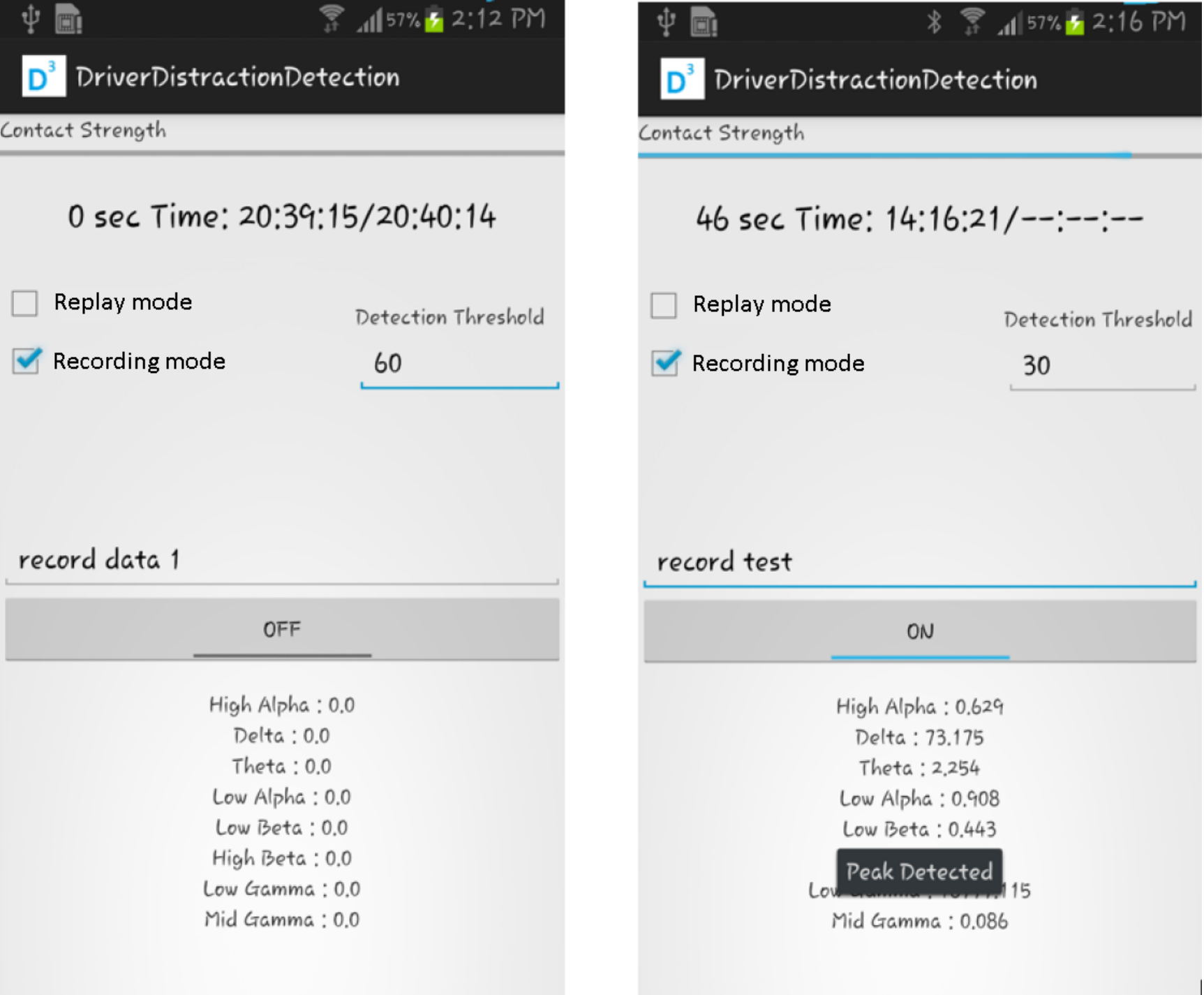}
  \caption{\textbf{An example screen from the Android application interface.} The application displays connection strength with the headband (top), the raw EEG signals, and power values of each frequency band (bottom). A voice prompt alerts the user when the peak detection algorithm detects a distraction event.}
  \label{fig:app1}
\end{figure}

Bluetooth protocol was used to establish the communication to the headband (Fig. \ref{fig:arch}). An Android API was used to retrieve the EEG data on the smartphone \cite{9}. The messages exchanged between the integrated chip in the headband, and API were parsed to obtain the necessary data on the phone. All the data was obtained at a frequency of 1 Hz except the raw data that was sampled at 512 Hz. Fig. \ref{fig:app1} shows our Android API, which retrieves the raw EEG data and the power plots of different frequency bands.

\begin{figure}[htb]
\centering
  \includegraphics[width=0.6\linewidth]{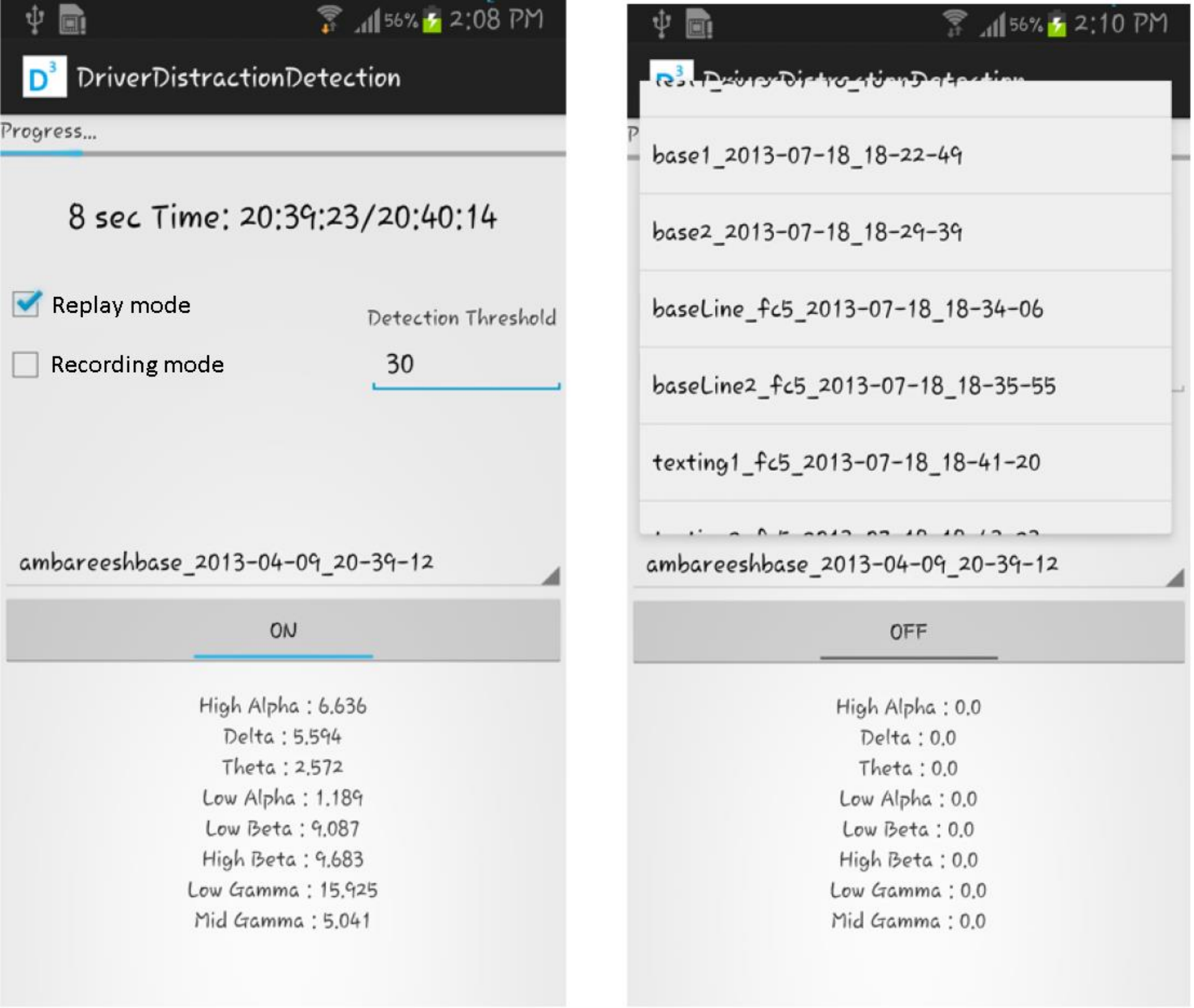}
  \caption{\textbf{An example screen from the Android application interface in the replay mode.} The replay mode is used to calibrate and test the peak detection algorithm for each subject separately to minimize false alerts.}
  \label{fig:app2}
\end{figure}

Our Brain Mobile Interface API implements a real-time capturing of EEG signals on the phone without any delay. The system consists of two modes: (1) Recording mode and (2) Replay mode. In the recording mode, the application shows the contact strength of the headband at the top to indicate a user the signal quality - a complete status bar for full strength. User records the EEG data via a toggle button on the screen.  All the calculated EEG parameters are made available to the user on the go. In the replay mode, the saved files in the phone memory can be retrieved for display and further analysis off the phone for driver behavior metrics (Fig. \ref{fig:app2}).

\subsection{\textbf{Mobile EEG Data Processing}}
\begin{figure}[h!]
\centering
  \includegraphics[width=0.8\linewidth]{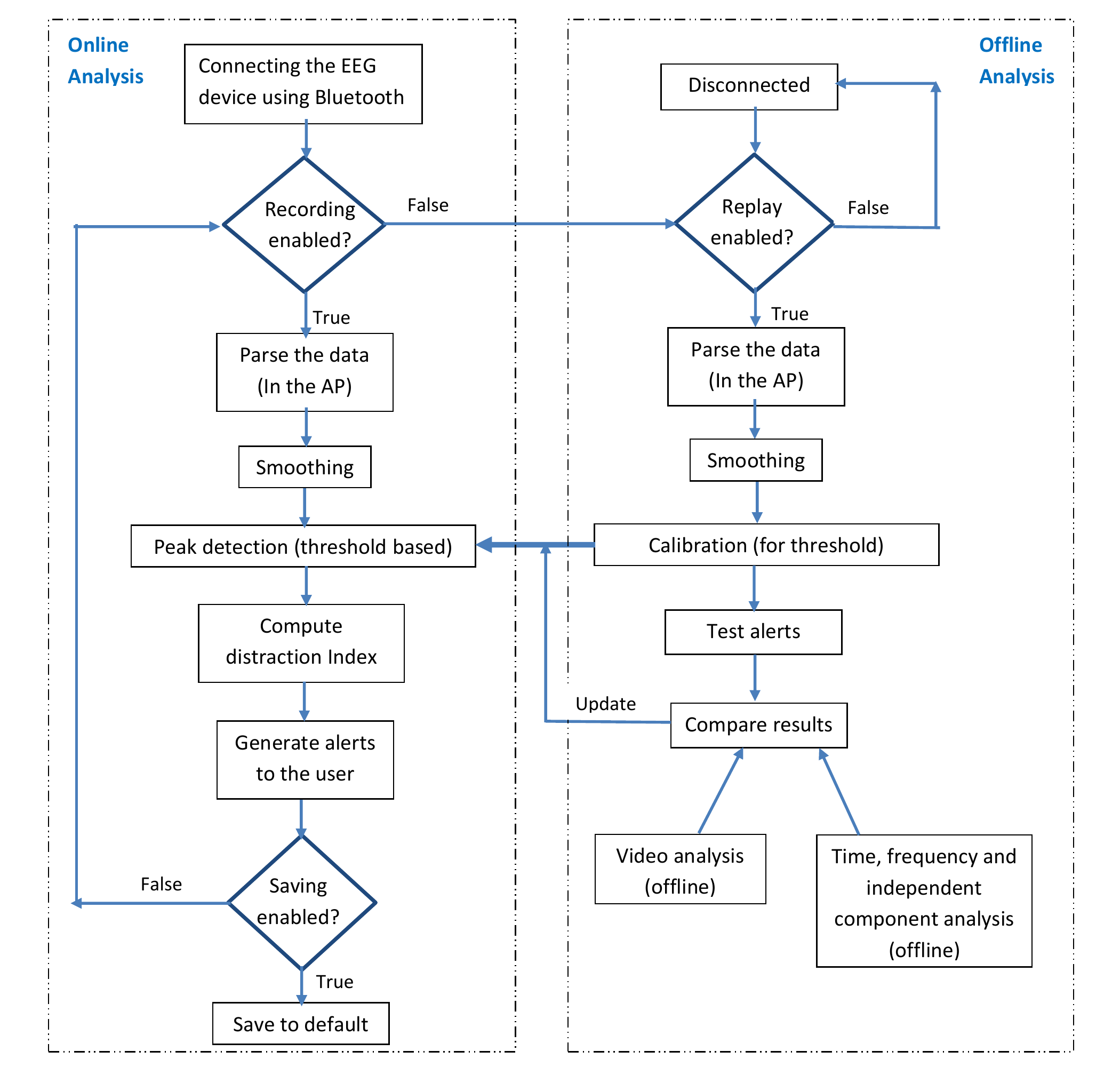}
  \caption{\textbf{Flowchart representing functionality of the system.}}
  \label{fig:flow}
\end{figure}

Distraction occurs in a high-low frequency pattern as observed in the power spectrums of the data from the subjects. Hence, a peak detection algorithm is useful at the start of a distraction event. Our detection algorithm was tailored to various customized thresholds for a subject and tested during the replay mode of the API (Fig. \ref{fig:flow}). The alerts were in the form voice prompts to the user if the power in frequency bands or a combination of frequency bands passed the customized threshold level. The instantaneous value of the Distraction Index was used to differentiate the extent of distraction.
\section{Conclusions and Future Work}
In sum, we have shown that using a single dry sensor in the frontal region of the scalp (FC5) is sufficient to detect driver distraction using an individuals' EEG signals in a mobile environment. We tested our mobile application interface for detecting distraction events using peaks in the higher frequency bands of EEG data. Our real-time detection and feedback can help to bring about the cultural change needed in distraction driving intuitively and create safer driving scenarios. The scientific understanding of distracted driving is necessary to guide policies and support laws to answer critical questions for investment in next car technologies.

More work is in progress to develop an active real-time application guide to assist drivers in situations of active or passive distraction using their EEG signals via smart phones, including the combination of atomic events detection such as swipe, tap and pinch.
In future, we plan to investigate the behavioral change of the drivers while interacting with our application. We are also working on identifying other driver behaviors such as taking turns, lane changes, sudden acceleration, braking time and reaction time using the brain signals. This multi-paradigm integration into a single mobile application will lead to predicting drivers' behavior based on their current state. It would be a substantial improvement in ensuring the health and safety of the drivers and other drivers who are becoming a part of the vehicle-to-vehicle communication network. The classification performance can be further improved using statistical feature selection algorithms. Classifier fusion methods \cite{Esmaeili} can also be used to extract EEG feature vectors that can further improve the detection capability of a distraction activity.

\section{Acknowledgements}
This work is partially supported by the National Science Foundation under grants CNS-0751205, and CNS-1229700. We thank Dr. Jason M. Pittman for proof-reading and providing constructive criticism of the manuscript.

\bibliographystyle{plain}
\bibliography{BibBajwaDantu}

\end{document}